\documentclass{bmcart}

\usepackage[utf8]{inputenc} 

\def\includegraphics{}

\startlocaldefs
\usepackage{amsmath,amssymb,epsfig,color, amsthm}
\usepackage{endnotes}
\let\footnote=\endnote

\newtheorem{theorem}{Theorem}
\newtheorem*{proof*}{Proof:}
\newtheorem{lemma}{Lemma}
\newtheorem{proposition}{Proposition}

\newtheorem*{remark*}{Remark:}

\endlocaldefs

\begin{document}

\begin{frontmatter}

\begin{fmbox}
\dochead{Research}


\title{One- and Two-Way Relay Optimization for MIMO Interference Networks}


\author[
   addressref={aff1},                   
   corref={aff1},                       
   email={m.khandaker@ucl.ac.uk}   
]{\inits{MRA}\fnm{Muhammad R A} \snm{Khandaker}}
\author[
   addressref={aff1},
   email={kai-kit.wong@ucl.ac.uk}
]{\inits{KK}\fnm{Kai-Kit} \snm{Wong}}


\address[id=aff1]{
  \orgname{Department of Electronic and Electrical Engineering, University College London}, 
  \street{Torrington Place},                 
  \city{London}                              
  \postcode{WC1E 7JE},                       
  \cny{UK}                                   
}



\end{fmbox}


\begin{abstractbox}

\begin{abstract} 
This paper considers multiple-input multiple-output (MIMO) relay communication in multi-cellular (interference) systems in which MIMO source-destination pairs communicate simultaneously. It is assumed that due to severe attenuation and/or shadowing effects, communication links can be established only with the aid of a relay node. The aim is to minimize the maximal mean-square-error (MSE) among all the receiving nodes under constrained source and relay transmit powers. Both one- and two-way amplify-and-forward (AF) relaying mechanisms are considered. Since the exactly optimal solution for this practically appealing problem is intractable, we first propose optimizing the source, relay, and receiver matrices in an alternating fashion. Then we contrive a simplified {\em semidefinite programming (SDP) solution} based on the error covariance matrix decomposition technique, avoiding the high complexity of the iterative process. Numerical results reveal the effectiveness of the proposed schemes.
\end{abstract}


\begin{keyword}
\kwd{interference}
\kwd{MIMO}
\kwd{two-way}
\kwd{relay}
\kwd{optimization}
\end{keyword}


\end{abstractbox}
%

\end{frontmatter}




\section{Introduction}
Due to scarcity of frequency spectrum in practical wireless networks, multiple communicating pairs are motivated to share a common time-frequency channel to ensure efficient use of the available spectrum. Co-channel interference (CCI) is, however, one of the main deteriorating factors in such networks that adversely affect the system performance. The impact is more obvious in 5G heterogeneous networks where there is oceanic volume of interference due to hyper-dense frequency reuse among small-cell and macro cell base stations. Therefore it is important to develop schemes to mitigate the CCI, which has been a major research direction in wireless communications over the past decades.

In the literature, various schemes have been proposed to control CCI at an acceptable level. A conventional approach in MIMO systems is to exploit spatial diversity for suppressing CCI \cite{tse_book, jrnl_duality, jrnl_mur1, icspcs10, conf_outage_mimo}. Such spatial diversity technique has been used to solve many power control problems in interference systems for different network setups. In \cite{jpcbf}, a power control scheme has been designed with receive diversity only, whereas joint transmit-receive beamforming has been considered in \cite{jpcbf, jtxrxd} for interference systems. However, the incorporation of the spatial diversity at the transmitter side in \cite{jtxrxd}, results in lower total transmit power compared to that in \cite{jpcbf}.

On the other hand, there is synergy between multiple antenna and relaying technologies. The latter is particularly useful to reestablish communications in case of a broken channel between source and destination. Hence relaying has been considered in interference networks in order to afford longer source-destination distance \cite{mp2p, mp2p_mants, jrnl_inter}. Both \cite{mp2p, mp2p_mants} considered network beamforming for minimizing total relay transmit power, whereas in \cite{jrnl_inter, apcc11inter}, an iterative transceiver optimization scheme has been proposed to minimize total source and relay transmit power.

While the works in \cite{jpcbf, jtxrxd, mp2p, mp2p_mants, jrnl_inter} all considered minimizing the total transmit power of interference networks, another important performance metric, which concerns more about the quality of communications, is the mean-square-error (MSE) for signal estimation \cite{apcc10dpc, sum_duality, conf_psr_spawc16}. In \cite{khoa_inter, khoa_inter_eura, khoa_inter_twc}, the sum minimum MSE (MMSE) was considered to design iterative algorithms for MIMO interference relay systems taking the direct links between the source and destination nodes into consideration, and in \cite{khoa_tway_int}, similar problem has been considered ignoring the direct links between the communicating parties. Nonetheless, the sum MMSE criterion runs the risk that some of the receivers may suffer from unacceptably high MSEs. Also, the works in \cite{khoa_inter, khoa_inter_eura, khoa_inter_twc} considered one-way relaying only.


Due to the increasing demands on multimedia applications, in particular, the notion of emerging wireless communications terminologies such as Big data, ultra-high spectral efficiency is essential in future wireless networks, including 5G, to provide ADSL-like user experience aspired by $2020$. The above-mentioned one-way relay systems suffer from a substantial performance loss in terms of spectral efficiency due to the pre-log factor of $1/2$ persuaded by the fact that two channel uses are required for each end-to-end transmission.

Two-way relay systems have hence been proposed to overcome the loss of spectral efficiency in such one-way relay methods \cite{tway_rate_reg, tway_mimo, tway_dist}. Utilizing the concept of analog network coding \cite{tway_dist}, communication in a two-way relay channel can be accomplished in two phases: the multiple access (MAC) phase and the broadcast (BC) phase. During the MAC phase, all the users simultaneously send their messages to an intermediate relay node, whereas in the BC phase, the relay retransmits the received information to the users. As each user knows its own transmitted signals, each user can cancel the self-interference and decode the intended message. The capacity region of multi-pair two-way relay networks in the deterministic channel was characterized in \cite{tway_cap_reg}. Later in \cite{tway_inter_dof}, the achievable total degrees of freedom in a two-way interference MIMO relay channel were also studied. Most recently in \cite{fd_wsrm}, the transceivers in a full-duplex MIMO interference system were optimized based on the weighted sum-rate maximization criterion.

In this paper, we consider a $K$-user MIMO interference system where each of the pairs can communicate only with the aid of a relay node thus ignoring the direct source-destination links. The direct links are understood to be in deep shadowing and hence negligible. Both one- and two-way amplify-and-forward (AF) relaying mechanisms are considered. All nodes are assumed to be equipped with multiple antennas so as to afford simultaneous transmission of multiple data streams. Our aim is to develop joint transceiver optimization algorithms for minimizing the worst-user MSE (min-max MSE)\footnote{The min-max MSE criterion is considered by many to be more desirable than the min-sum MSE criterion in \cite{khoa_inter, khoa_inter_eura, khoa_inter_twc} because fairness is imposed and weaker users are not being sacrificed for the minimization of the sum.} subject to the source and relay power constraints. It can be verified that the problem is strictly non-convex, and thus it is difficult to find an analytical solution. To tackle this, we first devise an algorithm to optimize the source, relay, and receiver matrices alternatingly by decomposing the original non-convex problem into convex subproblems. To avoid the complexity of the iterative process, we then extend the error covariance matrix decomposition technique applied to point-to-point MIMO relay systems in \cite{simpli_relay} to interference MIMO relay systems in this paper. More specifically, under practically reasonable high first-hop signal-to-noise ratio (SNR) assumption, we demonstrate that the problem can be decomposed into two standard semidefinite programming (SDP) problems to optimize source and relay matrices separately. Note that high SNR assumption has also been made in \cite{lim_feed} to simplify the joint codebook design problem in single-user MIMO relay systems and in \cite{jrnl_mcast, jrnl_mcast_mu} for multicasting MIMO relay design. Hence our work is a generalization to multi-pair communication scheme taking co-channel interference into account.

The remainder of this paper is lined-up as follows. In Section~\ref{sec_sys}, the interference MIMO relay system model is introduced. The joint optimal transmitter, relay, and receiver beamforming optimization schemes are developed in Section~\ref{sec_1way} and Section~\ref{sec_2way}, respectively, for one-way and two-way relaying. Section~\ref{sec_sim} provides simulation results to analyze the performance of the proposed algorithms in various system configurations before concluding remarks are made in Section~\ref{sec_con}.

\section{System Model}\label{sec_sys}%
Let us consider a communication scenario, as illustrated in Fig. \ref{sysmod}, where each of the $K$ source nodes communicates with the corresponding destination node sharing the same frequency channel via a common relay node. The direct link between each transmitter-receiver pair is assumed to be broken due to strong attenuation and/or shadowing effects. The $k$th source, the relay, and the $k$th destination nodes are assumed to be equipped with $N_{{\rm s},k}$, $N_{{\rm r}}$, and $N_{{\rm d},k}$ antennas, respectively.

\section{One-Way Relaying}\label{sec_1way}
In this section, we consider that communication takes place in one direction only. The relay node is assumed to work in half-duplex mode which implies that the actual communication between the source and destination nodes is accomplished in two time slots. In the first time slot, the source nodes transmit the linearly precoded signal vectors ${\bf B}_k{\bf s}_k, k = 1, \cdots, K,$ to the relay node. The received signal vector at the relay node is therefore given by
\begin{equation}
{\bf y}_{\rm r} = \sum_{k=1}^K{\bf H}_{k}{\bf B}_k{\bf s}_k + {\bf
n}_{\rm r}, \label{yr}
\end{equation}
where ${\bf H}_{k}$ denotes the $N_{{\rm r}}\times N_{{\rm s},k}$ Gaussian channel matrix between the $k$th source node and the intermediate relay node, ${\bf s}_k$ is the $N_{{\rm b},k} \times 1\, (1 \leq N_{{\rm b},k} \leq N_{{\rm s},k})$ transmit symbol vector with covariance ${\bf I}_{N_{\rm b},k}$, ${\bf B}_k$ is the $N_{{\rm s},k}\times N_{{\rm b},k}$ source precoding matrix, and ${\bf n}_{{\rm r}}$ is the $N_{{\rm r}}\times 1$ additive white Gaussian noise (AWGN) vector introduced at the relay node. Let us denote $N_{{\rm b}} = \sum_{k=1}^KN_{{\rm b},k}$ as the total number of data streams transmitted by all the source nodes. In order to successfully transmit $N_{\rm b}$ independent data streams simultaneously through the relay, the relay node must be equipped with $N_{\rm r}\geq N_{\rm b}$ antennas.

After receiving ${\bf y}_{\rm r}$, the relay node simply multiplies the signal vector by an $N_{{\rm r}}\times N_{{\rm r}}$ precoding matrix ${\bf F}$ and transmits the amplified version of ${\bf y}_{\rm r}$ in the second time slot. Thus the relay's $N_{{\rm r}}\times 1$ transmit signal vector ${\bf x}_{{\rm r}}$ is given by
\begin{equation}
{\bf x}_{\rm r} = {\bf F}{\bf y}_{\rm r}.\label{xr}
\end{equation}
Accordingly, the signal received at the $k$th destination node can be expressed as
\begin{align}
{\bf y}_{{\rm d},k}&={\bf G}_{k} {\bf x}_{{\rm r}} + {\bf n}_{{\rm d},k}\nonumber\\
&= \underbrace{{\bf G}_{k}{\bf F}{\bf H}_{k}{\bf B}_k{\bf s}_k}_{\text{desired signal}} + \underbrace{{\bf G}_{k}{\bf F} \sum_{j=1\atop j\neq k}^K{\bf H}_{j}{\bf B}_j{\bf s}_j}_{\text{interference signal}} + \underbrace{{\bf G}_{k}{\bf F}{\bf
n}_{\rm r} + {\bf n}_{{\rm d},k}}_{\text{noise}},\label{ydl}\\
&=\bar{\bf H}_{k}{\bf s}_k + \bar{\bf n}_{{\rm d},k}, ~\mbox{for }k=1,\dots,K,
\end{align}
where ${\bf G}_{k}$ denotes the $N_{{\rm d},k}\times N_{{\rm r}}$ complex channel matrix between the relay node and the $k$th destination node, ${\bf n}_{{\rm d},k}$ is the $N_{{\rm d},k}\times 1$ AWGN vector introduced at the $k$th destination node, $\bar{\bf H}_{k}\triangleq {\bf G}_{k}{\bf F}{\bf H}_{k}{\bf B}_k$ is the equivalent source-destination channel matrix, and $\bar{\bf n}_{{\rm d},k}\triangleq {\bf G}_{k}{\bf F} (\sum_{j=1\atop j\neq k}^K{\bf H}_{j}{\bf B}_j{\bf s}_j + {\bf n}_{\rm r}) + {\bf n}_{{\rm d},k}$ is the equivalent noise vector. All noises are assumed to be independent and identically distributed (i.i.d.) complex Gaussian random variables with mean zero and variance $\sigma_{\rm n}^2$, where ${\rm n} \in \{{\rm r}, {\rm d}\}$ indicates the noise introduced at the relay or at the destination.

\begin{remark*}
Note that the interference term in \eqref{ydl} does not appear in the received signal of the single-user MIMO relay system considered in \cite{lim_feed} or in the multicasting MIMO relay system considered in \cite{jrnl_mcast, jrnl_mcast_mu}. Hence the subsequent analyses remain considerably simpler in \cite{lim_feed, jrnl_mcast, jrnl_mcast_mu}, whereas we need to deal with this troublesome interference term in this paper.
\end{remark*}

Considering the input-output relationship at the relay node given in (\ref{xr}), the average transmit power consumed by the MIMO relay node is defined as
\begin{equation}
{\rm tr}\big({\rm E}\{{\bf x}_{{\rm r}} {\bf x}_{{\rm r}}^H\}\big)= {\rm tr}\big({\bf F} {\boldsymbol \Psi}{\bf F}^H\big),\label{Prk}
\end{equation}
where ${\rm tr}(\cdot)$ denotes trace of a matrix, ${\rm E}\{\cdot\}$ indicates statistical expectation, and ${\boldsymbol \Psi}\triangleq {\rm E}\{{\bf y}_{{\rm r}}{\bf y}_{{\rm r}}^H\} =\sum_{k=1}^K{\bf H}_{k}{\bf B}_k{\bf B}_k^H{\bf H}_{k}^H +\sigma_{\rm r}^2{\bf I}_{N_{{\rm r}}}$ represents the covariance matrix of the signal vector received at the relay node.

For signal detection, linear receivers are used at the destination nodes for simplicity reasons. Denoting ${\bf W}_k$ as the $N_{{\rm d},k}\times N_{{\rm b},k}$ receiver matrix used by the $k$th destination node, the corresponding estimated signal vector ${\hat{\bf s}}_k$ can be written as
\begin{equation}\label{shat}
{\hat{\bf s}}_k = {\bf W}_k^H{\bf y}_{{\rm d},k}, ~~\mbox{for }k = 1, \dots, K,
\end{equation}
where $(\cdot)^H$ indicates the conjugate transpose (Hermitian) of a matrix (vector). Thus the MSE of signal estimation at the $k$th receiver can be expressed as
\begin{align}
E_k &={\rm tr}\left({\bf E}_k \triangleq {\rm E}\left[(\hat{\bf s}_k-{\bf s}_k) (\hat{\bf
s}_k - {\bf s}_k)^{H}\right]\right),\nonumber\\
&= {\rm tr} \left(\begin{array}{r}
{\bf I}_{N_{\rm b},k} - {\bf W}_k^H{\bf G}_{k}{\bf F}{\bf H}_{k}{\bf B}_k - {\bf B}_k^H{\bf H}_{k}^H{\bf F}^H{\bf G}_k^H{\bf W}_k\\
+ \sum_{j=1}^K{\bf W}_k^H{\bf G}_{k}{\bf F}{\bf H}_j{\bf B}_j{\bf B}_j^H{\bf H}_j^H{\bf F}^H{\bf G}_k^H{\bf W}_k\\
+ \sigma_{\rm r}^2{\bf W}_k^H{\bf G}_{k}{\bf F}{\bf F}^H{\bf G}_k^H{\bf W}_k + \sigma_{\rm d}^2{\bf W}_k^H{\bf W}_k
\end{array}\right),\nonumber\\
&={\rm tr} \left(({\bf W}_k^H {\bar{\bf H}}_k- {\bf I}_{N_{{\rm b},k}})({\bf W}_k^H {\bar{\bf H}}_k- {\bf I}_{N_{{\rm b},k}})^H + {\bf W}_k^H {\bar{\bf C}}_k{\bf W}_k\right),\nonumber\\
& \qquad\qquad\qquad\qquad\qquad\qquad\qquad\qquad\qquad ~~\mbox{for } k = 1, \dots, K,\label{ek}
\end{align}
where ${\bf E}_k$ denotes the error covariance matrix at the $k$th receiver, and
\begin{equation}
\bar {\bf C}_k\triangleq \sum_{j=1\atop j\ne k}^K{\bf G}_{k}{\bf F}{\bf H}_j{\bf B}_j{\bf B}_j^H{\bf H}_j^H{\bf F}^H{\bf G}_k^H + \sigma_{\rm r}^2{\bf G}_{k}{\bf F}{\bf F}^H{\bf G}_k^H + \sigma_{\rm d}^2{\bf I}_{N_{\rm d}}.
\end{equation}
is the combined interference and noise covariance matrix.

In the following subsections, we develop optimization approaches that minimize the worst-user MSE among all the receivers subject to source and relay power constraints.

\subsection{Problem Formulation}
In this section, we formulate the joint source and relay precoding optimization problem for MIMO interference systems.
Our aim is to minimize the maximal MSE among all the source-destination pairs yet satisfying the transmit power constraints at the source as well as the relay nodes. To fulfill this aim, the following joint optimization problem is formulated:
\begin{subequations}
\label{minmax0}
\begin{eqnarray}
\min_{\{{\bf B}_k\}, {\bf F}, \{{\bf W}_k\}} \!\!\!& &\!\!\! \max_k\;\; E_k \label{mnx_a}\\
{\rm s.t.} \!\!\!& &\!\!\! {\rm tr}\left({\bf F} {\boldsymbol \Psi} {\bf F}^H\right)\leq P_{{\rm r}} \label{mnx_a_c1}\\
\!\!\!& &\!\!\! {\rm tr}({\bf B}_k{\bf B}_k^H)\leq P_{{\rm s},k},~~\mbox{for }k = 1, \dots, K \label{mnx_a_c2}
\end{eqnarray}
\end{subequations}
where (\ref{mnx_a_c1}) and (\ref{mnx_a_c2}), respectively, constrains the transmit power at the relay node and the $k$th transmitter to $P_{{\rm r}} > 0$, $P_{{\rm s}, k} > 0$. Our next endeavour is to develop optimal solutions for this problem. Note that the problem is strictly non-convex with matrix variables appearing in quadratic form, and hence any closed-form solution is intractable. Therefore, we first resort to developing an iterative algorithm for the problem and then propose a sub-optimal solution which has lower computational complexity.

\subsection{Iterative Joint Transceiver Optimization}
In this subsection, we investigate the non-convex source, relay, and destination filter design problem in an alternating fashion. We tend to optimize one group of variables while fixing the others. Given source and relay matrices $\{{\bf B}_k\}$, ${\bf F}$, the optimal receiver matrices $\{{\bf W}_k\}$ are obtained through solving the unconstrained optimization problem of $\min_{{\bf W}_k} E_k$, since $E_k$ does not depend on ${\bf W}_j$, for $j\neq k$, and ${\bf W}_k$ does not appear in constraints (\ref{mnx_a_c1}) and (\ref{mnx_a_c2}). Using the matrix derivative formulas, the gradient $\nabla_{{\bf W}_k^H}\left({\rm tr}\left({\bf E}_k\right)\right)$ can be written as
\begin{multline}\label{dek}
\nabla_{{\bf W}_k^H}\left({\rm tr}\left({\bf E}_k\right)\right)= - {\bf G}_{k}{\bf F}{\bf H}_{k}{\bf B}_k + \sum_{j=1}^K{\bf G}_{k}{\bf F}{\bf H}_j{\bf B}_j{\bf B}_j^H{\bf H}_j^H{\bf F}^H{\bf G}_k^H{\bf W}_k\\
+ \sigma_{\rm r}^2{\bf G}_{k}{\bf F}{\bf F}^H{\bf G}_k^H{\bf W}_k
+ \sigma_{\rm d}^2{\bf W}_k, ~~\mbox{for }k = 1, \dots, K.
\end{multline}
Equating $\nabla_{{\bf W}_k^H}\left({\rm tr}\left({\bf E}_k\right)\right) = {\bf 0}$ yields the linear MMSE receive filter given by 
\begin{multline}
{\bf W}_k = \left(\sum_{j=1}^K{\bf G}_{k}{\bf F}{\bf H}_j{\bf B}_j{\bf B}_j^H{\bf H}_j^H{\bf F}^H{\bf G}_k^H + \sigma_{\rm r}^2{\bf G}_{k}{\bf F}{\bf F}^H{\bf G}_k^H + \sigma_{\rm d}^2{\bf I}_{N_{{\rm d},k}}\right)^{-1}\\
\times{\bf G}_{k}{\bf F}{\bf H}_{k}{\bf B}_k \label{Wk}
\end{multline}
where $(\cdot)^{-1}$ indicates the inversion operation of a matrix.

Then for given source and receiver matrices $\{{\bf B}_k\}$ and $\{{\bf W}_k\}$, the relay precoding matrix ${\bf F}$ optimization problem can be formulated as
\begin{subequations}
\label{minmaxF}
\begin{eqnarray}
\min_{{\bf F}} \!\!\!& &\!\!\! \max_k\;\; E_k \label{mnxF_o}\\
{\rm s.t.} \!\!\!& &\!\!\! {\rm tr}\left({\bf F} {\boldsymbol \Psi} {\bf F}^H\right)\leq P_{{\rm r}}.\label{mnxF_c1}
\end{eqnarray}
\end{subequations}
Note that \eqref{minmaxF} is non-convex with a matrix variable since ${\bf F}$ appears in quadratic form in the objective function as well as in the constraint. However, we can reformulate this problem as an SDP using Schur complement \cite{mat_ana} as follows. By introducing a matrix ${\boldsymbol \Xi}_k$ we conclude from the second equation in \eqref{ek} that the $k$-th link MSE will be upper-bounded if
\begin{equation}
- {\bf W}_k^H{\bf G}_{k}{\bf F}{\bf H}_{k}{\bf B}_k - {\bf B}_k^H{\bf H}_{k}^H{\bf F}^H{\bf G}_k^H{\bf W}_k + {\bf W}_k^H{\bf G}_{k}{\bf F}{\boldsymbol \Psi}{\bf F}^H{\bf G}_k^H{\bf W}_k  \preceq {\boldsymbol \Xi}_k.
\end{equation}
In the above inequality, ${\bf A} \preceq {\bf B}$ indicates that the matrix ${\bf B} - {\bf A}$ is positive semidefinite (PSD). Now, by introducing a matrix ${\boldsymbol \Phi}$ such that ${\bf F} {\boldsymbol \Psi} {\bf F}^H\preceq {\boldsymbol \Phi},$ and a scaler variable $\tau_{\rm r}$, the relay optimization problem \eqref{minmaxF} can be transformed to
\begin{subequations}
\label{minmaxF2}
\begin{eqnarray}
\min_{\tau_{\rm r}, {\bf F}, \{{\boldsymbol \Xi}_k\}, {\boldsymbol \Phi}} \!\!\!& &\!\!\! \tau_{\rm r} \label{mnxF2_o}\\
{\rm s.t.} \!\!\!& &\!\!\! {\rm tr}\left({\boldsymbol \Xi}_k\right) + {\rm tr}\left({\bf W}_k^H{\bf W}_k\right)  + N_{{\rm b},k}\le \tau_{\rm r}, ~~\mbox{for }k = 1, \dots, K, \label{mnxF2_c1}\\
\!\!\!& &\!\!\! \left[\begin{array}{cc}
{\boldsymbol \Xi}_k + {\bf W}_k^H{\bf G}_{k}{\bf F}{\bf H}_{k}{\bf B}_k + {\bf B}_k^H{\bf H}_{k}^H{\bf F}^H{\bf G}_k^H{\bf W}_k & {\bf W}_k^H{\bf G}_{k}{\bf F}\\
{\bf F}^H{\bf G}_k^H{\bf W}_k & {\boldsymbol \Psi}^{-1}
\end{array}\right]\succeq {\bf 0},\nonumber\\
\!\!\!& &\!\!\! \qquad\qquad\qquad\qquad\qquad\qquad ~~\mbox{for }k = 1, \dots, K,\label{mnxF2_c2}\\
\!\!\!& &\!\!\! \left[\begin{array}{cc}
{\boldsymbol \Phi} & {\bf F}\\
{\bf F}^H & {\boldsymbol \Psi}^{-1}
\end{array}\right]\succeq {\bf 0}\label{mnxF2_c3}\\
\!\!\!& &\!\!\! {\rm tr}\left({\boldsymbol \Phi}\right) \le P_{\rm r} \label{mnxF2_c4}
\end{eqnarray}
\end{subequations}
where we have used the Schur complement to obtain \eqref{mnxF2_c2} and \eqref{mnxF2_c3}. Note that the problem \eqref{minmaxF2} is an SDP problem which is convex and can, as a result, be efficiently solved using interior-point based solvers \cite{cvx} at a maximal complexity order of ${\mathcal O}\big((K + 2N_{\rm r}^2 + \sum_{k=1}^{K}N_{{\rm b},k}^2 + 2)^{3.5}\big)$ \cite{ipm}. However, the actual complexity is usually much less in many practical cases. Interested readers are referred to \cite{ipm} for a detailed analysis of the computational complexity based on interior-point methods.

Finally, we optimize the source matrices $\{{\bf B}_k\}$ using the relay matrix ${\bf F}$ and the receiver matrices $\{{\bf W}_k\}$ known from the previous steps. Let us define $\tilde{\bf H}_{k,j} \triangleq {\bf W}_k^H{\bf G}_{k}{\bf F}{\bf H}_j$. Applying the matrix identity ${\rm vec}({\bf ABC})=({\bf C}^T \!\! \otimes \! {\bf A}) {\rm vec}({\bf B})$, we can rewrite $E_k$ in \eqref{ek} as
\begin{align}
E_k &= \sum_{j=1}^K {\bf b}_j^H ({\bf I}_{N_{{\rm b},j}} \!\!\otimes \!(\tilde{\bf H}_{k,j}^H\tilde{\bf H}_{k,j})) {\bf b}_j - \big({\rm vec}(\tilde{\bf H}_{k,j})\big)^T{\bf b}_k  - {\bf b}_k^H{\rm vec}(\tilde{\bf H}_{k,j}^H) + \theta_k,\label{Ei}
\end{align}
where the vector ${\bf b}_k \triangleq {\rm vec}({\bf B}_k)$ is created by stacking all the columns of the matrix ${\bf B}_k$ on top of each other, $\theta_k \triangleq {\rm tr}(\sigma_{\rm r}^2{\bf W}_k^H{\bf G}_{k}{\bf F}{\bf F}^H{\bf G}_k^H{\bf W}_k + \sigma_{\rm d}^2{\bf W}_k^H{\bf W}_k)+ N_{{\rm b},k}$, and $\otimes$ indicates matrix Kronecker product. Let us now denote
\begin{equation}
\left\{\begin{aligned}
\tilde{\bf G}_k &\triangleq {\rm bd}\big({\bf I}_{N_{{\rm b},1}} \!\! \otimes \! (\tilde{\bf H}_{k,1}^H\tilde{\bf H}_{k,1}), \dots,{\bf I}_{N_{{\rm b},K}} \!\! \otimes \! (\tilde{\bf H}_{k,K}^H\tilde{\bf H}_{k,K})\big),\\
{\bf c}_k &\triangleq \big[({\rm vec}(\tilde{\bf C}_{k,1}))^T, \dots, ({\rm vec}(\tilde{\bf C}_{k,K}))^T\big]^T,\\
{\bf b} &\triangleq \big[{\bf b}_1^T, \dots, {\bf b}_K^T\big]^T,
\end{aligned}\right.
\end{equation}
where ${\rm bd}(\cdot)$ constructs a block-diagonal matrix taking the parameter matrices as the diagonal blocks, $\tilde{\bf C}_{k,k} = \tilde{\bf H}_{k,k}$ and $\tilde{\bf C}_{k,j} = {\bf 0}_{N_{{\rm b},k}\times N_{{\rm s},j}}$, if $j \ne k$. The MSE in \eqref{Ei} can be rewritten as
\begin{equation}\label{Ei2}
E_k ={\bf b}^H \tilde{\bf G}_k{\bf b}- {\bf c}_k^H {\bf b}- {\bf b}^H {\bf c}_k+\theta_k.
\end{equation}
By introducing ${\bf M}_{k} \triangleq {\bf F}{\bf H}_{k}$, the power constraints in \eqref{mnx_a_c1} can be rewritten as
\begin{equation}\label{powr}
{\bf b}^H {\bf M} {\bf b}\leq \bar{P}_{{\rm r}}, ~~\mbox{for }k=1,\dots,K,
\end{equation}
where ${\bf M} \triangleq {\rm bd}\big({\bf I}_{N_{{\rm b},1}} \!\! \otimes \! ({\bf M}_{1}^H{\bf M}_{1}), \dots,{\bf I}_{N_{{\rm b},K}} \!\! \otimes \! ({\bf M}_{K}^H{\bf M}_{K})\big)$, and $\bar{P}_{{\rm r}}\!=\! P_{{\rm r}}\!-\!\sigma_{\rm r}^2{\rm tr}({\bf F}{\bf F}^H)$. Using \eqref{Ei2} and \eqref{powr}, problem \eqref{minmax0} can be written as
\begin{subequations}
\label{qcqp_itb}
\begin{eqnarray}
\min_{\bf b}\!\!\!& &\!\!\! \max_k \;\; {\bf b}^H \tilde{\bf G}_k{\bf b}- {\bf c}_k^H {\bf b}- {\bf b}^H {\bf c}_k+\theta_k \label{opt1_1}\\
{\rm s.t.} \!\!\!& &\!\!\! {\bf b}^H {\bf M} {\bf b}\leq \bar{P}_{{\rm r}} \label{opt1_2}\\
\!\!\!& &\!\!\! {\bf b}^H \mathcal{I}_k {\bf b} \leq P_{{\rm s},k}, ~~\mbox{for }k=1,\dots,K,\label{opt1_3}
\end{eqnarray}
\end{subequations}
where $\mathcal{I} \triangleq {\rm bd}(\mathcal{I}_{k1}, \dots, \mathcal{I}_{kk}, \dots, \mathcal{I}_{kK})$ with $\mathcal{I}_{kk}={\bf I}_{N_{{\rm s},k}N_{{\rm b},k}}$ and $\mathcal{I}_{kj}={\bf 0}$, if $j\neq k$. Problem \eqref{qcqp_itb} is a standard quadratically-constrained quadratic program (QCQP) which can be solved using off-the-shelf convex optimization toolboxes \cite{cvx}. In the following, we also provide an SDP formulation of problem \eqref{qcqp_itb}:

\begin{subequations}\label{sdp_itb}
\begin{eqnarray}
\min_{t_{\rm s},{\bf b}} \!\!\!& &\!\!\! \tau_{\rm s} \label{opt2_1}\\
{\rm s.t.} \!\!\!& &\!\!\! \left(\!
                             \begin{array}{cc}
                               \tau_{\rm s}-\theta_k+ {\bf c}_k^H {\bf b}+ {\bf b}^H {\bf c}_k  & {\bf b}^H \\
                               {\bf b} & \tilde{\bf G}_k^{-1} \\
                             \end{array}\!\!
                           \right) \succcurlyeq {\bf 0}, ~~\mbox{for }k=1,\dots,K,\label{opt2_2}\\
\!\!\!& &\!\!\! \left(\!
                  \begin{array}{cc}
                    \bar{P}_{{\rm r}} & {\bf b}^H \\
                    {\bf b} & {\bf M}^{-1} \\
                  \end{array}\!\!
                \right) \succcurlyeq {\bf 0},\label{opt2_3}\\
\!\!\!& &\!\!\! \left(\!
                  \begin{array}{cc}
                    P_{{\rm s},k} & {\bf b}^H\mathcal{I}_k^\frac{1}{2} \\
                    \mathcal{I}_k^\frac{1}{2}{\bf b} &  {\bf I}_p\\
                  \end{array}\!\!
                \right) \succcurlyeq {\bf 0}, ~~\mbox{for }k=1,\dots,K,\label{opt2_4}
\end{eqnarray}
\end{subequations}
where $\tau_{\rm s}$ is a slack variable and $p \triangleq \sum_{k=1}^K N_{{\rm s},k} N_{{\rm b},k}$. The problem \eqref{sdp_itb} can be solved at a maximal complexity order of ${\mathcal O}\big((\sum_{k=1}^{K}N_{{\rm b},k}^2 + 1)^{3.5}\big)$ \cite{ipm}. The proposed iterative optimization technique for solving the original problem \eqref{minmax0} is summarized in Table~\ref{tab_it}.


Since in each step of the iterative algorithm we solve a convex subproblem to update one set of variables, the conditional update of each set will either decrease or maintain the objective function (\ref{mnx_a}). From this observation, a monotonic convergence of the iterative algorithm follows. However, the overall computational complexity of the iterative algorithm increases as the multiple of the number of iterations required until convergence. Thus the complexity of the iterative algorithms is often reasonably high. Note that the sum-MSE based iterative algorithms proposed in \cite{khoa_inter, khoa_inter_eura, khoa_inter_twc} have similar complexity orders. Hence in the following subsection, we contrive an algorithm for the joint optimization problem such that the computational overhead is substantially reduced.

\subsection{Simplified Joint Optimization Algorithm}
In the previous subsection, we optimized the source, relay, and receiver matrices in an alternating fashion. Here, we propose a simplified approach to solve problem \eqref{minmax0} using the error covariance matrix decomposition technique. The following theorem paves the foundation of the simplified algorithm.

\begin{theorem}\label{thm_struc}
For given $\{{\bf B}_k\}$ and $\{{\bf W}_k\}$, the optimum relaying matrix ${\bf F}$ for minimizing the worst-user MSE has the form:
\begin{equation}
{\bf F} = \sum_{k = 1}^K{\bf T}_k{\bf D}_k^H = {\bf T}{\bf D}^H,
\end{equation}
where ${\bf T}\triangleq \left[{\bf T}_1, \dots, {\bf T}_K\right]$ and ${\bf D}\triangleq \left[{\bf D}_1, \dots, {\bf D}_K\right]$ with ${\bf T}_k$ and ${\bf D}_k$, respectively, defined as
\begin{equation}
{\bf T}_k \triangleq \lambda_{{\rm e},k}\left(\sum_{i=1}^K\lambda_{{\rm e},i}{\bf G}_i^H{\bf W}_i{\bf W}_i^H{\bf G}_i + \lambda_{\rm r}{\bf I}_{N_{\rm r}}\right)^{-1}{\bf G}_{k}^H{\bf W}_k
\end{equation}
and
\begin{equation}
{\bf D}_k \triangleq \left(\sum_{j=1}^K{\bf H}_j{\bf B}_j{\bf B}_j^H{\bf H}_j^H + \sigma_{\rm r}^2{\bf I}_{N_{\rm r}}\right)^{-1}{\bf H}_{k}{\bf B}_k, \label{Dk}
\end{equation}
$\lambda_{\rm r}$ and $\lambda_{{\rm e},k}, \forall k,$ are the corresponding Lagrange multipliers as defined in Appendix~\ref{proof_thm_struc}.
\end{theorem}

\begin{proof*}
See Appendix~\ref{proof_thm_struc}.
\end{proof*}

Note that ${\bf D}_k = \left({\bf H}_k{\bf B}_k{\bf B}_k^H{\bf H}_k^H + \sum_{j=1\atop j\neq k}^K{\bf H}_j{\bf B}_j{\bf B}_j^H{\bf H}_j^H + \sigma_{\rm r}^2{\bf I}_{N_{\rm r}}\right)^{-1}{\bf H}_{k}{\bf B}_k$ can be regarded as the MMSE receive filter of the first-hop MIMO channel for the $k$th transmitter's signal received at the relay node given by \eqref{yr}.

The implication of the structure of the relay amplifying matrix in the proposed simplified design can be observed while applying the following theorem.

\begin{theorem}\label{thm_mse}
The MSE term appearing in \eqref{mnx_a} can be equivalently decomposed into
\begin{multline}\label{mse_dec}
E_k = {\rm tr}\left({\bf I}_{N_{\rm b},k} + {\bf B}_k^H{\bf H}_{k}^H{\boldsymbol\Psi}_{\bar k}^{-1}{\bf H}_k{\bf B}_k\right)^{-1}\\
 + {\rm tr} \left(\left({\bf B}_k^H{\bf H}_{k}^H{\boldsymbol\Psi}^{-1}{\bf H}_k{\bf B}_k\right)^{-1} + \tilde{\bf T}^H{\bf G}_k^H{\bf G}_{k}\tilde{\bf T}\right)^{-1},
\end{multline}
where ${\boldsymbol\Psi}_{\bar k} \triangleq {\boldsymbol\Psi} - {\bf H}_{k}{\bf B}_k{\bf B}_k^H{\bf H}_{k}^H = \sum_{j=1\atop j\neq k}^K{\bf H}_{j}{\bf B}_j{\bf B}_j^H{\bf H}_{j}^H + \sigma_{\rm n}^2{\bf I}_{N_{{\rm r}}}$ and $\tilde{\bf T}$ is defined in Appendix~\ref{proof_thm_mse}.
\end{theorem}

\begin{proof*}
See Appendix~\ref{proof_thm_mse}.
\end{proof*}

Even given the structure, an analytical optimal solution to the joint optimization problem is still difficult to obtain due to the cross-link interference from the relay node to the destination nodes. Therefore, we resort to develop an efficient suboptimal solution. The following proposition provides the foundation of the proposed simplified suboptimal solution.

\begin{proposition}\label{prop_approx}
In the practically reasonably high SNR regime, the term ${\bf B}_k^H{\bf H}_{k}^H\\ \times{\boldsymbol\Psi}^{-1}{\bf H}_k{\bf B}_k$ in \eqref{mse_dec} can be approximated as ${\bf B}_k^H{\bf H}_{k}^H{\boldsymbol\Psi}^{-1}{\bf H}_k{\bf B}_k \approx {\bf I}_{N_{{\rm b},k}}$.
\end{proposition}

\begin{proof*}
See Appendix~\ref{proof_prop_approx}.
\end{proof*}

The result in {\em Proposition~\ref{prop_approx}} is guided by the observation that the eigenvalues of ${\bf B}_k^H{\bf H}_{k}^H{\boldsymbol\Psi}^{-1}{\bf H}_k{\bf B}_k$ approach unity with increasing first-hop SNR.
It will be demonstrated in Section~\ref{sec_sim} through numerical simulations that such an approximation results in negligible performance loss while reducing the computational complexity significantly. Applying {\em Proposition~\ref{prop_approx}}, the transmit power of the relay node defined in \eqref{Prk} can be expressed as ${\rm tr}\left({\bf F} {\boldsymbol \Psi} {\bf F}^H\right) = {\rm tr}(\tilde{\bf T}{\bf B}_k^H{\bf H}_{k}^H{\boldsymbol\Psi}^{-1}{\bf H}_k{\bf B}_k\tilde{\bf T}^H) = {\rm tr}(\tilde{\bf T}\tilde{\bf T}^H)$. Therefore, problem \eqref{minmax0} can be approximated as
\begin{subequations}
\label{minmax2}
\begin{eqnarray}
\min_{\{{\bf B}_k\}, \{{\bf W}_k\}, \tilde{\bf T}} \!\!\!& &\!\!\! \max_k \;\; {\rm tr}\left({\bf I}_{N_{\rm b},k} + {\bf B}_k^H{\bf H}_{k}^H{\boldsymbol\Psi}_{\bar k}^{-1}{\bf H}_k{\bf B}_k\right)^{-1}\nonumber\\
\!\!\!& &\!\!\! \qquad\qquad\qquad\qquad\qquad + {\rm tr} \left({\bf I}_{N_{{\rm b},k}} + \tilde{\bf T}^H{\bf G}_k^H{\bf G}_{k}\tilde{\bf T}\right)^{-1} \label{mnx2_o}\\
{\rm s.t.} \!\!\!& &\!\!\! {\rm tr}({\bf B}_k{\bf B}_k^H)\leq P_{{\rm s},k},~~\mbox{for }k = 1, \dots, K,\label{mnx2_c2}\\
\!\!\!& &\!\!\! {\rm tr}\left(\tilde{\bf T}\tilde{\bf T}^H\right)\leq P_{\rm r}. \label{mnx2_c1}
\end{eqnarray}
\end{subequations}
Note that the optimal receiver matrices $\{{\bf W}_k\}$ can be obtained as in \eqref{Wk}. Interestingly, the source and relay optimization variables $\{{\bf B}_k\}$ and $\tilde{\bf T}$ are separable both in the objective function as well as in the constraints in problem \eqref{minmax2}.
Therefore, applying the results from {\em Theorem~\ref{thm_mse}} and {\em Proposition~\ref{prop_approx}}, we can decompose the problem (\ref{minmax2}) into the following source precoding matrices optimization problem:
\begin{subequations}
\label{decs}
\begin{eqnarray}
\min_{\{{\bf B}_k\}} \!\!\!& &\!\!\! \max_k ~{\rm tr}\left({\bf I}_{N_{\rm b},k} + {\bf B}_k^H{\bf H}_{k}^H{\boldsymbol\Psi}_{\bar k}^{-1}{\bf H}_k{\bf B}_k\right)^{-1} \label{decs_o}\\
{\rm s.t.} \!\!\!& &\!\!\! {\rm tr}({\bf B}_k{\bf B}_k^H)\leq P_{{\rm s},k},~~\mbox{for }k = 1, \dots, K,\label{decs_c}
\end{eqnarray}
\end{subequations}
and the relay amplifying matrix optimization problem:
\begin{subequations}
\label{decrL}
\begin{eqnarray}
\min_{\tilde{\bf T}} \!\!\!& &\!\!\! \max_k \quad {\rm tr}\left(\left[{\bf I}_{N_{{\rm b},k}} + \tilde{\bf T}^H{\bf G}_{k}^H{\bf G}_{k}\tilde{\bf T}\right]^{-1}\right) \label{decr_oL}\\
{\rm s.t.} \!\!\!& &\!\!\! {\rm tr}(\tilde{\bf T}\tilde{\bf T}^H)\leq P_{{\rm r}}. \label{decr_cL}
\end{eqnarray}
\end{subequations}

Note that the objective function in \eqref{decs_o} can be interpreted as the MSE of the $k$th transmitter's signal vector ${\bf s}_k$. In particular, the equivalent received signal for the $k$th transmitter's signal in the first hop received at the relay node is given by ${\bf y}_{\rm r}^{(k)} = {\bf H}_{k}{\bf B}_k{\bf s}_k + \sum_{j \ne k}^K{\bf H}_{j}{\bf B}_j{\bf s}_j + {\bf n}_{\rm r}$, treating other users' signals as noise. As such, the corresponding MMSE receiver is given by ${\bf D}_{k}$ in \eqref{Dk}. Thus the MSE expression in \eqref{decs_o} actually represents the equivalent first-hop MSE of the $k$th transmitter's signal ${\bf s}_k$. Given the corresponding MMSE receiver ${\bf D}_{k}$, \eqref{decs_o} can be rewritten as
\begin{eqnarray}
E_{{\rm s},k} \!\!\!&\triangleq&\!\!\! {\rm tr}\left({\bf D}_{k}^H\left({\boldsymbol\Psi} + \sigma_{\rm r}^2{\bf I}_{N_{\rm r}}\right){\bf D}_{k} - {\bf D}_{k}^H{\bf H}_k{\bf B}_k - {\bf B}_k^H{\bf H}_k^H{\bf D}_{k} + {\bf I}_{N_{\rm b},k}\right) \nonumber\\
\!\!\!&=&\!\!\! {\rm tr}\left(\left({\bf D}_{k}^H{\bf H}\boldsymbol\Upsilon_{k}{\bf B} - \boldsymbol\Omega_k\right)\left({\bf D}_{k}^H{\bf H}\boldsymbol\Upsilon_{k}{\bf B} - \boldsymbol\Omega_k\right)^H + \sigma_{\rm r}^2{\bf D}_{k}^H{\bf D}_{k}\right) \nonumber\\
\!\!\!&=&\!\!\! \left\|{\rm vec}\left({\bf D}_{k}^H{\bf H}\boldsymbol\Upsilon_{k}{\bf B} - \boldsymbol\Omega_k\right)\right\|_2^2 + \sigma_{\rm r}^2{\rm tr}\left({\bf D}_k^H{\bf D}_{k}\right) \nonumber\\
\!\!\!&=&\!\!\! \left\|\left[\begin{array}{c}
\omega_k\\
\left({\bf I}_{N_{\rm r}}\otimes{\bf D}_{k}^H{\bf H}\boldsymbol\Upsilon_{k}\right){\rm vec}\left({\bf B}\right) - {\rm vec}\left(\boldsymbol\Omega_k\right)
\end{array}\right]\right\|_2^2,
\end{eqnarray}
where $\omega_k \triangleq \sigma_{\rm r}\sqrt{{\rm tr}({\bf D}_{k}^H{\bf D}_{k})}$ and $\boldsymbol\Upsilon_k \triangleq \left[\boldsymbol\Upsilon_{k1}, \dots, \boldsymbol\Upsilon_{kk}, \dots, \boldsymbol\Upsilon_{kK}\right]$ with $\boldsymbol\Upsilon_{kk}={\bf I}_{N_{\rm r}}$ and $\boldsymbol\Upsilon_{kj}={\bf 0}$, if $j\neq k$. Introducing an auxiliary variable $t_{\rm s}$, problem \eqref{decs} can be rewritten as the following second-order cone program (SOCP):
\begin{subequations}
\label{decs2}
\begin{eqnarray}
\min_{\{{\bf B}_k\}, t_{\rm s}} \!\!\!& &\!\!\! t_{\rm s} \label{decs2_o}\\
{\rm s.t.} \!\!\!& &\!\!\! \!\!\left\|\left[\!\!\begin{array}{c}
\omega_k\\
\left({\bf I}_{N_{\rm r}}\otimes{\bf D}_{k}^H{\bf H}\boldsymbol\Upsilon_{k}\right){\rm vec}\left({\bf B}\right) - {\rm vec}\left(\boldsymbol\Omega_k\right)\end{array}\!\!\right]\right\|_2 \!\!\le\! t_{\rm s},\mbox{for }k = 1, \dots, K,\label{decs2_c1}\\
\!\!\!& &\!\!\! \!\!\left\|{\rm vec}\left({\bf B}_k\right)\right\|_2 \le \sqrt{P_{{\rm s},k}},~~\mbox{for }k = 1, \dots, K,\label{decs2_c2}
\end{eqnarray}
\end{subequations}
which can be efficiently solved by standard optimization packages at a complexity order of ${\mathcal O}\big((\sum_{k=1}^{K}N_{{\rm b},k}^2 + 1)^{3}\big)$ \cite{ipm}. Thus, we can update $\{{\bf D}_{k}\}$ and $\{{\bf B}_{k}\}$ in an alternating fashion.

Regarding the relay amplifying matrix optimization, by introducing $\tilde{\bf T}^H\tilde{\bf T} \triangleq {\bf Q}$, the relay matrix optimization problem (\ref{decrL}) can be equivalently transformed to
\begin{subequations}
\label{decr2}
\begin{eqnarray}
\min_{{\bf Q}\succeq 0} \!\!\!& &\!\!\! \max_k \quad{\rm tr}\!\left(\left[{\bf I}_{N_{\rm d},k} + {\bf G}_k{\bf Q}{\bf G}_k^H\right]^{-1}\right) + N_{{\rm b},k} - N_{{\rm d},k} \label{decr_o2}\\
{\rm s.t.} \!\!\!& &\!\!\! {\rm tr}({\bf Q})\leq P_{{\rm r}}.
\end{eqnarray}
\end{subequations}
Let us now introduce a matrix variable ${\bf Y}_k\succeq \left({\bf I}_{N_{\rm d},k} + {\bf G}_k{\bf Q}{\bf G}_k^H\right)^{-1}$, and a scalar variable $t_{\rm r}$. Using these variables, the relay optimization problem (\ref{decr2}) can be equivalently rewritten as the following SDP:
\begin{subequations}
\label{decr3}
\begin{eqnarray}
\min_{t_{\rm r}, {\bf Q}, \{{\bf Y}_k\}} \!\!\!& &\!\!\! t_{\rm r} \label{decr_o3}\\
{\rm s.t.} \!\!\!& &\!\!\! {\rm tr}({\bf Y}_k)\leq t_{\rm r},~~\mbox{for }k = 1, \dots, K,\\
\!\!\!& &\!\!\! {\rm tr}({\bf Q})\leq P_{{\rm r}},\label{decr_c13}\\
\!\!\!& &\!\!\!   \left(\begin{array}{cc}
                                       {\bf Y}_k & {\bf I}_{N_{\rm d},k},\\
                                       {\bf I}_{N_{\rm d},k} & {\bf I}_{N_{\rm d},k} + {\bf G}_k{\bf Q}{\bf G}_k^H \\
                                     \end{array}\!\!
                                   \right) \succeq 0, ~~\mbox{for } k = 1,\dots, K,\label{decr_c23}\\
\!\!\!& &\!\!\! t_{\rm r}\geq 0,\\
\!\!\!& &\!\!\! {\bf Q}\succeq 0.\label{decr_c33}
\end{eqnarray}
\end{subequations}
Problem \eqref{decr3} is convex and the globally optimal solution can be easily obtained \cite{cvx}. The complexity order of solving problem \eqref{decr3} is at most ${\mathcal O}\big((\sum_{k=1}^{K}N_{{\rm b},k}^2 + \sum_{k=1}^{K}N_{{\rm d},k}^2 + K + 2)^{3.5}\big)$ \cite{ipm}. Note that in the simplified algorithm, only the source matrices are obtained in an alternating fashion.
The overall joint optimization procedure is summarized in Table~\ref{tab_simpli}.




\section{Two-Way Relaying}\label{sec_2way}
Two-way relaying is being considered as a promising technique for future generation wireless systems since two-way relaying can significantly improve spectral efficiency. Hence in this section, we consider two-way relaying in an interference MIMO relay system where each pair of users transmit signals to each other through the assisting relay node. The information exchange in the two-way relay channel is accomplished in two time slots: MAC phase and the BC phase. During the MAC phase, all the users simultaneously send their messages to the relay node. Thus the signal vector received at the relay node during the MAC phase can be expressed as
\begin{equation}\label{yr2w}
{\bf y}_{\rm r} = \sum_{k=1}^{2K}{\bf H}_{k}{\bf B}_k{\bf s}_k + {\bf n}_{\rm r},
\end{equation}
where ${\bf H}_{K+k} \triangleq {\bf G}_{k}^T$ for $k = 1, \dots, K$ and ${\bf n}_{{\rm r}}$ is the $N_{{\rm r}}\times 1$ AWGN vector received at the relay node.

Upon receiving ${\bf y}_{\rm r}$, the relay node linearly precodes the signal vector by an $N_{{\rm r}}\times N_{{\rm r}}$ amplifying matrix ${\bf F}$ and transmits the $N_{{\rm r}}\times 1$ precoded signal vector ${\bf x}_{{\rm r}}$ in the MAC phase:
\begin{equation}
{\bf x}_{\rm r} = {\bf F}{\bf y}_{\rm r}.\label{xr2w}
\end{equation}
The received signal at the $k$th user in the BC phase is given by
\begin{eqnarray}
{\bf y}_{k}\!\!\!\!&=&\!\!\!\! {\bf H}_{k}^T {\bf x}_{{\rm r}} + {\bf n}_{{\rm d},k}\nonumber\\
\!\!\!\!&=&\!\!\!\! {\bf H}_{k}^T{\bf F}{\bf H}_{\bar k}{\bf B}_{\bar k}{\bf s}_{\bar k} + {\bf H}_{k}^T{\bf F} \!\!\left(\sum_{j=1\atop j\neq {\bar k}}^{2K}{\bf H}_{j}{\bf B}_j{\bf s}_j + {\bf
n}_{\rm r}\!\!\right)\!\! + {\bf n}_{{\rm d},k},\mbox{for }k=1, \dots, 2K,\label{ydl2w}
\end{eqnarray}
where we have defined $\bar k$ as the index of user $k$'s partner (e.g., ${\bar 1} = K+1, \overline{K+1} = 1$), ${\bf n}_{{\rm d},k}$ is the $N_{{\rm d},k}\times 1$ AWGN vector at the $k$th destination node. As in the case of the one-way relaying system, all noises are assumed to be i.i.d.~complex Gaussian random variables with mean zero and variance $\sigma_{\rm n}^2$.

Since the transmitting node $k$ knows its own signal vector ${\bf s}_k$ and the full CSI of the corresponding source-destination link ${\bf H}_{k}^T{\bf F}{\bf H}_{k}{\bf B}_{k}$, each transmitter can completely cancel the self-interference component in \eqref{ydl2w}. Thus, the effective received signal vector at the $k$th receiving node is given by
\begin{eqnarray}
{\bf y}_{k}\!\!\!\!&=&\!\!\!\! {\bf H}_{k}^T{\bf F}{\bf H}_{\bar k}{\bf B}_{\bar k}{\bf s}_{\bar k} + {\bf H}_{k}^T{\bf F} \left(\sum_{ j\neq k, {\bar k}}^{2K}{\bf H}_{j}{\bf B}_j{\bf s}_j + {\bf n}_{\rm r}\right) + {\bf n}_{{\rm d},k},\label{ydl2w2}\\
\!\!\!\!&=&\!\!\!\! \bar{\bf H}_{k}{\bf s}_{\bar k} + \bar{\bf n}_{{\rm d},k}, ~~\mbox{for }k=1, \dots, 2K.
\end{eqnarray}

Using (\ref{xr2w}), the transmission power required at the relay node can be defined as
\begin{equation}
{\rm tr}\big({\rm E}\{{\bf x}_{{\rm r}} {\bf x}_{{\rm r}}^H\}\big)= {\rm tr}\big({\bf F} {\boldsymbol \Psi}{\bf F}^H\big),\label{Prk2w}
\end{equation}
where ${\boldsymbol \Psi} \triangleq {\rm E}\{{\bf y}_{{\rm r}}{\bf y}_{{\rm r}}^H\} = \sum_{k=1}^{2K}{\bf H}_{k}{\bf B}_k{\bf B}_k^H{\bf H}_{k}^H + \sigma_{\rm r}^2{\bf I}_{N_{{\rm r}}}$ is the covariance matrix of the signal received at the relay node from all the transmitters. Furthermore, the MSE of the estimated signal using an $N_{\rm d}\times N_{\rm b}$ linear weight matrix ${\bf W}_k$ at the $k$th receiving node can be expressed as
\begin{equation}\label{ek2w}
E_k = {\rm tr} \left(\begin{array}{r}
{\bf I}_{N_{\rm s},k} - {\bf W}_k^H{\bf H}_{k}^T{\bf F}{\bf H}_{\bar k}{\bf B}_{\bar k} - {\bf B}_{\bar k}^H{\bf H}_{\bar k}^H{\bf F}^H{\bf H}_k^*{\bf W}_k\\
+ \sum_{j=1\atop j\ne k}^{2K}{\bf W}_k^H{\bf H}_{k}^T{\bf F}{\bf H}_j{\bf B}_j{\bf B}_j^H{\bf H}_j^H{\bf F}^H{\bf H}_{k}^*{\bf W}_k\\
+ \sigma_{\rm r}^2{\bf W}_k^H{\bf H}_{k}^T{\bf F}{\bf F}^H{\bf H}_{k}^*{\bf W}_k + \sigma_{\rm d}^2{\bf W}_k^H{\bf W}_k
\end{array}\right),\mbox{for }k = 1, \dots, 2K.
\end{equation}
Similar to the case of one-way relaying, the problem of optimizing the transmit, relay, and receive matrices for the two-way scenario can be formulated as
\begin{subequations}
\label{minmax02}
\begin{eqnarray}
\min_{\{{\bf B}_k\}, {\bf F}, \{{\bf W}_k\}} \!\!\!& &\!\!\! \max_k\;\; E_k \label{mnx2_a}\\
{\rm s.t.} \!\!\!& &\!\!\! {\rm tr}\left({\bf F} {\boldsymbol \Psi} {\bf F}^H\right)\leq P_{{\rm r}} \label{mnx2_a_c1}\\
\!\!\!& &\!\!\! {\rm tr}({\bf B}_k{\bf B}_k^H)\leq P_{{\rm s},k},~~\mbox{for }k = 1, \dots, 2K,\label{mnx2_a_c2}
\end{eqnarray}
\end{subequations}
where (\ref{mnx2_a_c1}) and (\ref{mnx2_a_c2}) indicates the corresponding transmit power constraints.

\subsection{Iterative Joint Transceiver Optimization}
Similar to the one-way relaying scenario, it can be shown that the transmitter, relay, and receiver matrices can be optimized in an alternating fashion through solving convex sub-problems. In each iteration of the algorithm, the receiver weight matrices are updated as follows:
\begin{multline}\label{Wk2w}
{\bf W}_k = \left(\sum_{j=1\atop j \ne k}^{2K}\!\!{\bf G}_{k}{\bf F}{\bf H}_j{\bf B}_j{\bf B}_j^H{\bf H}_j^H{\bf F}^H{\bf G}_k^H + \sigma_{\rm r}^2{\bf G}_{k}{\bf F}{\bf F}^H{\bf G}_k^H + \sigma_{\rm d}^2{\bf I}_{N_{\rm d}}\right)^{-1}\\
\times{\bf G}_{k}{\bf F}{\bf H}_{k}{\bf B}_k, ~~\mbox{for } k = 1, \dots, 2K.
\end{multline}
The relay beamforming matrix ${\bf F}$ is optimized through solving the following SDP problem:
\begin{subequations}
\label{minmaxF2w}
\begin{eqnarray}
\min_{\tau_{\rm r}, {\bf F}, \{{\boldsymbol \Xi}_k\}, {\boldsymbol \Phi}} \!\!\!& &\!\!\! \tau_{\rm r} \label{mnxF2w_o}\\
{\rm s.t.} \!\!\!& &\!\!\! {\rm tr}\left({\boldsymbol \Xi}_k\right) + {\rm tr}\left({\bf F}_k^H{\bf W}_k\right) \le \tau_{\rm r},\label{mnxF2w_c1}\\
\!\!\!& &\!\!\! \left[\begin{array}{cc}
{\boldsymbol \Xi}_k + {\bf W}_k^H{\bf H}_{k}^T{\bf F}{\bf H}_{\bar k}{\bf B}_{\bar k} + {\bf B}_{\bar k}^H{\bf H}_{\bar k}^H{\bf F}^H{\bf H}_k^*{\bf W}_k & {\bf W}_k^H{\bf H}_{k}^T{\bf F}\\
{\bf F}^H{\bf H}_{k}^*{\bf W}_k & {\boldsymbol \Psi}_{\bar k}^{-1}
\end{array}\right]\succeq {\bf 0}, \nonumber\\
\!\!\!& &\!\!\!\qquad\qquad\qquad\qquad\qquad\qquad\qquad\qquad \mbox{for }k=1,\dots, 2K,\label{mnxF2w_c2}\\
\!\!\!& &\!\!\! \left[\begin{array}{cc}
{\boldsymbol \Phi} & {\bf F}\\
{\bf F}^H & {\boldsymbol \Psi}^{-1}
\end{array}\right]\succeq {\bf 0},\label{mnxF2w_c3}\\
\!\!\!& &\!\!\! {\rm tr}\left({\boldsymbol \Phi}\right) \le P_{\rm r},\label{mnxF2w_c4}
\end{eqnarray}
\end{subequations}
where we have defined
\begin{equation}
\left\{\begin{aligned}
{\bf F} {\boldsymbol \Psi} {\bf F}^H&\preceq {\boldsymbol \Phi},\\
-{\bf W}_k^H{\bf H}_{k}^T{\bf F}{\bf H}_{\bar k}{\bf B}_{\bar k} - {\bf B}_{\bar k}^H{\bf H}_{\bar k}^H{\bf F}^H{\bf H}_k^*{\bf W}_k + {\bf W}_k^H{\bf H}_{k}^T{\bf F}{\boldsymbol \Psi}_{\bar k}{\bf F}^H{\bf H}_{k}^*{\bf W}_k  &\preceq {\boldsymbol \Xi}_k.
\end{aligned}\right.
\end{equation}
Finally, the optimal source precoding matrices are obtained by solving
\begin{subequations}
\label{sdp_itb2w}
\begin{eqnarray}
\min_{t_{\rm s},{\bf b}} \!\!\!& &\!\!\! \tau_{\rm s} \label{opt2w_1}\\
{\rm s.t.} \!\!\!& &\!\!\! \left(\!
                             \begin{array}{cc}
                               \tau_{\rm s}-\theta_k+ {\bf c}_k^H {\bf b}+ {\bf b}^H {\bf c}_k  & {\bf b}^H \\
                               {\bf b} & \tilde{\bf G}_k^{-1} \\
                             \end{array}\!\!
                           \right) \succcurlyeq {\bf 0},~~\mbox{for }k=1,\dots, 2K,\label{opt2w_2}\\
\!\!\!& &\!\!\! \left(\!
                  \begin{array}{cc}
                    \bar{P}_{{\rm r}} & {\bf b}^H \\
                    {\bf b} & {\bf M}^{-1} \\
                  \end{array}\!\!
                \right) \succcurlyeq {\bf 0},\label{opt2w_3}\\
\!\!\!& &\!\!\! \left(\!
                  \begin{array}{cc}
                    P_{{\rm s},k} & {\bf b}^H\mathcal{I}_k^\frac{1}{2} \\
                    \mathcal{I}_k^\frac{1}{2}{\bf b} &  {\bf I}_p\\
                  \end{array}\!\!
                \right) \succcurlyeq {\bf 0},~~\mbox{for }k=1,\dots, 2K,\label{opt2w_4}
\end{eqnarray}
\end{subequations}
where
\begin{subequations}
\begin{align}
\theta_k &\triangleq {\rm tr}(\sigma_{\rm r}^2{\bf W}_k^H{\bf G}_{k}{\bf F}{\bf F}^H{\bf G}_k^H{\bf W}_k + \sigma_{\rm d}^2{\bf W}_k^H{\bf W}_k)+ N_{{\rm b},k}, ~~\mbox{for }k=1,\dots, 2K,\\
\tilde{\bf G}_k &\triangleq {\rm bd}\big({\bf I}_{N_{{\rm b},1}} \!\! \otimes \! (\tilde{\bf H}_{k,1}^H\tilde{\bf H}_{k,1}), \cdots,{\bf I}_{N_{{\rm b},2K}} \!\! \otimes \! (\tilde{\bf H}_{k,2K}^H\tilde{\bf H}_{k,2K})\big),\mbox{for } k=1,\dots, 2K,\\
{\bf c}_k &\triangleq \big[({\rm vec}(\tilde{\bf C}_{k,1}))^T, \dots, ({\rm vec}(\tilde{\bf C}_{k,2K}))^T\big]^T,\\
\tilde{\bf C}_{k,k} &= \tilde{\bf H}_{k,k},\\
\tilde{\bf C}_{k,j}& = {\bf 0}_{N_{{\rm b},k}\times N_{{\rm s},j}}, ~~\mbox{for }j \ne k,\\
{\bf b} &\triangleq  \big[{\bf b}_1^T, \dots, {\bf b}_{2K}^T\big]^T,\\
{\bf M} &\triangleq {\rm bd}\big({\bf I}_{N_{{\rm b},1}} \!\! \otimes \! ({\bf M}_{1}^H{\bf M}_{1}), \dots,{\bf I}_{N_{{\rm b},K}} \!\! \otimes \! ({\bf M}_{2K}^H{\bf M}_{2K})\big),\\
p &\triangleq \sum_{k=1}^{2K} N_{{\rm s},k} N_{{\rm b},k}.
\end{align}
\end{subequations}

\subsection{Simplified Non-Iterative Approach}
Assuming moderate SNR in the MAC phase, it can be shown, similar to the one-way relaying case, that the generic structure of the relay matrix ${\bf F}$ is defined as ${\bf F} = {\bf T}{\bf D}^H$. Using this particular structure of ${\bf F}$, the MSE at the $k$th receiver can be equivalently decomposed into two parts as shown below:
\begin{multline}\label{mse_dec_2w}
E_k ={\rm tr}\left({\bf I}_{N_{\rm b},k} + {\bf B}_k^H{\bf H}_{k}^H{\boldsymbol\Psi}_{\bar k}^{-1}{\bf H}_k{\bf B}_k\right)^{-1}\\
+ {\rm tr} \left(\left({\bf B}_k^H{\bf H}_{k}^H{\boldsymbol\Psi}_{\bar k}^{-1}{\bf H}_k{\bf B}_k\right)^{-1} + \tilde{\bf T}^H{\bf H}_{\bar k}^*{\bf H}_{\bar k}^T\tilde{\bf T}\right)^{-1}.
\end{multline}
Accordingly, the joint precoding design problem (\ref{minmax2}) can be decomposed into two sub-problems, namely, the source precoding matrices optimization problem:
\begin{subequations}
\label{decs_2w}
\begin{eqnarray}
\min_{\{{\bf B}_k\}} \!\!\!& &\!\!\! \max_k ~{\rm tr}\left({\bf I}_{N_{\rm b},k} + {\bf B}_k^H{\bf H}_{k}^H{\boldsymbol\Psi}_{\bar k}^{-1}{\bf H}_k{\bf B}_k\right)^{-1} \label{decs2w_o}\\
{\rm s.t.} \!\!\!& &\!\!\! {\rm tr}({\bf B}_k{\bf B}_k^H)\leq P_{{\rm s},k},~~\mbox{for }k = 1, \dots, K,\label{decs2w_c}
\end{eqnarray}
\end{subequations}
and the relay beamforming matrix optimization problem:
\begin{subequations}
\label{decrL2w}
\begin{eqnarray}
\min_{\tilde{\bf T}} \!\!\!& &\!\!\! \max_k \quad {\rm tr}\left(\left[{\bf I}_{N_{{\rm b},k}} + \tilde{\bf T}^H{\bf H}_{\bar k}^*{\bf H}_{\bar k}^T\tilde{\bf T}\right]^{-1}\right) \label{decr2w_oL}\\
{\rm s.t.} \!\!\!& &\!\!\! {\rm tr}(\tilde{\bf T}\tilde{\bf T}^H)\leq P_{{\rm r}},\label{decr2w_cL}
\end{eqnarray}
\end{subequations}
which can be solved following the similar approach as for the one-way relaying scenario.

\section{Numerical Simulations}\label{sec_sim}
In this section, we analyze the performance of the proposed one- and two-way MIMO relay interference system optimization algorithms through numerical examples. For simplicity, we assume that the source and the destination nodes are equipped with $N_{\rm s}$ and $N_{\rm d}$ antennas each, respectively, and $P_{{\rm s},k} = P_{\rm s},~\forall k$. We simulated a flat Rayleigh fading environment such that the channel matrices have zero-mean entries with variances $1/N_{\rm s}$ for ${\bf H}_{k},~\forall k$, and $1/N_{\rm r}$ for ${\bf G}_{k},~\forall k$. All the simulation results were obtained by averaging over $500$ independent channel realizations.

The performance of the proposed min-max MSE algorithms have been compared with that of the naive AF (NAF) algorithm in terms of both MSE and bit error rate (BER). The NAF algorithm is a simple baseline scheme that forwards the signals at the transmitters and the relay node assigning equal power to each data stream. In particular, the source and the relay matrices, in their simplest forms, in the NAF scheme are defined as
\begin{equation}
\left\{\begin{aligned}
{\bf B}_k& = \sqrt{P_{\rm s}/N_{\rm s}}\,{\bf I}_{N_{\rm s}},~~\mbox{for }k = 1, \dots, K,\\
{\bf F} &= \sqrt{P_{\rm r}/{\rm tr}({\boldsymbol \Psi})}\, {\bf I}_{N_{\rm r}}.
\end{aligned}\right.
\end{equation}

In the first example, we compare the performance of the proposed min-max MSE-based one-way algorithms with that of the sum-MSE minimization algorithm in \cite{khoa_inter} as well as the NAF approach in terms of the MSE normalized by the number of data streams (NMSE) with $K = 3$, $N_{\rm s} = 3, N_{\rm r} = 9$, and $N_{\rm d} = 3$. Fig.~\ref{mseL4s} shows the NMSE performance of the algorithms versus transmit power $P_{\rm s}$ with fixed $P_{\rm r} = 20$ dB. Note that for the proposed simplified {\em non-iterative} algorithm, we plot the NMSE of the user with the worst channel (Worst) as well as the average per-stream MSE of all the users (Avg.). On the other hand, for the rest of the algorithms, the worst-user NMSE has been plotted. The results clearly indicate that the proposed joint optimization algorithms consistently yield better performance compared to the existing schemes. It can also be revealed that the proposed iterative algorithm has the best MSE performance compared to the other approaches over the entire $P_{\rm s}$ range. It is no surprise that the NAF algorithm yields much higher MSE compared to the other schemes since the NAF algorithm performs no optimization operation. Most importantly, the iterative sum-MSE minimization algorithm in \cite{khoa_inter} always penalizes the user with the worst channel condition.

Since the NAF algorithm does not allocate the transmit power optimally, and equally divides the power among multiple data streams instead, the inter-stream interference and the inter-user interference increase significantly at higher transmit power. Hence the MSE of the NAF algorithm does not improve notably at higher transmit power.

Further analysis of the results in Fig.~\ref{mseL4s} reveals that the proposed simplified algorithm yields the worst-user MSE performance which is comaprable to that of the iterative algorithm, even at low $P_{\rm s}$ region. This observation illustrates that the approximation made in the simplified algorithm encounters negligible performance loss compared to the iterative optimal design. On the other hand, the computational complexity of the proposed simplified optimization is less than that of even one iteration of the iterative design, making it much more attractive for practical interference MIMO relay systems. The number of iterations required for convergence up to $10^{-3}$ in terms of MSE in a random channel realization for the iterative algorithm are listed in Table~\ref{tab2}.

In the next example, we focus on the proposed simplified optimization scheme and compare its performance with that of the proposed iterative approach and the NAF algorithm in terms of BER. Quadrature phase-shift keying (QPSK) signal constellations were assumed to modulate the transmitted signals and maximum-likelihood detection is applied at the receivers. We set $K = 3$, $N_{\rm s} = 2, N_{\rm r} = 6$, $N_{\rm d} = 3$, and transmit $1000N_{\rm s}$ randomly generated bits from each transmitter in each channel realization. The BER performance of the algorithms are shown in Fig.~\ref{berL2s} versus $P_{\rm s}$ with $P_{\rm r} = 20$dB. As we can see, the proposed simplified algorithm yields a much lower BER compared to the conventional NAF scheme.
Compared with the iterative approach the simplified algorithm has much lower computational task at the cost of marginal performance loss.

In the last couple of examples, we analyze the performance of the two-way MIMO relaying scheme. The NMSE performance of the two-way relaying algorithms is shown for different number of communication links $K$ in Fig.~\ref{2way_diffKs}. This time we set $N_{\rm s} = 2, N_{\rm r} = K N_{\rm s}$, and $N_{\rm d} = 6$ to plot the NMSE of the proposed algorithms versus $P_{\rm s}$ with $P_{\rm r} = 20$ dB. It can be clearly seen from Fig.~\ref{2way_diffKs} that as the number of links increases, the worst-user MSE keeps increasing. This is due to the additional cross-link interferences generated by the increased number of active users.

In Fig.~\ref{2way_diffAlgs}, the BER performance of the proposed two-way relaying algorithms has been compared with the sum-MSE based algorithms originally proposed for one-way relaying in \cite{khoa_inter, khoa_inter_eura, khoa_inter_twc}. QPSK signal constellations were assumed to modulate the transmitted signals. We set $N_{\rm s} = 2, K = 3, N_{\rm r} = K N_{\rm s}$, $N_{\rm d} = 6$, $P_{\rm r} = 20$dB, and transmit $1000N_{\rm s}$ randomly generated bits from each transmitter in each channel realization. Most importantly, the iterative sum-MSE minimization algorithms in \cite{khoa_inter, khoa_inter_eura, khoa_inter_twc} always penalize the user with the worst channel condition in the two-way relaying system.

\section{Conclusions}\label{sec_con}
We considered a two-hop interference MIMO relay system and developed schemes to minimize the worst-user MSE of signal estimation for both one- and two-way relaying schemes. At first, we proposed an iterative solution for both relaying schemes by solving several convex subproblems alternatingly and in an iterative fashion. Then to reduce the computational overhead of the optimization approach, we develop a simplified non-iterative algorithm using the error covariance matrix decomposition technique based on the high SNR assumption. Simulation results have illustrated that the proposed simplified approach performs nearly as well as the iterative approach, while offering significant reduction in computational complexity.

\theendnotes

\appendix
\section*{Appendices}
\addcontentsline{toc}{section}{Appendices}
\renewcommand{\thesubsection}{\Alph{subsection}}

\subsection{Proof of Theorem~\ref{thm_struc}}\label{proof_thm_struc}
For given $\{{\bf B}_k\}$ and $\{{\bf W}_k\}$, problem \eqref{minmax0} reduces to
\begin{subequations}
\label{minmaxP}
\begin{eqnarray}
\min_{{\bf F}} \!\!\!& &\!\!\! \tau \label{mnxP_o}\\
{\rm s.t.} \!\!\!& &\!\!\! E_k \le \tau \label{mnxP_c0},~~\mbox{for }k = 1, \dots, K,\\ 
\!\!\!& &\!\!\! {\rm tr}\left({\bf F} {\boldsymbol \Psi} {\bf F}^H\right)\leq P_{\rm r}. \label{mnxP_c1}
\end{eqnarray}
\end{subequations}
The Lagrangian function of problem \eqref{minmaxP} can be written as
\begin{multline}\label{Lek}
{\mathcal L}\left({\bf F}, \{\lambda_{{\rm s},k}\}, \lambda_{\rm r}\right)
=\tau + \sum_{k=1}^{K} \lambda_{{\rm e},k}{\rm tr} \!\!\left(\!\!\!
\begin{array}{r}
{\bf I}_{N_{\rm s},k} - 2{\rm Re}\left({\bf B}_k^H{\bf H}_{k}^H{\bf F}^H{\bf G}_k^H{\bf W}_k\right)\\
+ \sum_{j=1}^K{\bf W}_k^H{\bf G}_{k}{\bf F}{\bf H}_j{\bf B}_j{\bf B}_j^H{\bf H}_j^H{\bf F}^H{\bf G}_k^H{\bf W}_k\\
+ \sigma_{\rm r}^2{\bf W}_k^H{\bf G}_{k}{\bf F}{\bf F}^H{\bf G}_k^H{\bf W}_k + \sigma_{\rm d}^2{\bf W}_k^H{\bf W}_k - \tau
\end{array} \!\!\!\right)\\
+ \lambda_{\rm r}\left({\rm tr}\left({\bf F} \left(\sum_{k=1}^K{\bf H}_{k}{\bf B}_k{\bf B}_k^H{\bf H}_{k}^H + \sigma_{\rm r}^2{\bf I}_{N_{\rm r}}\right) {\bf F}^H\right) - P_{\rm r} \right).
\end{multline}
The derivative of the Lagrangian function over ${\bf F}^H$ is given by
\begin{multline}\label{dLek}
\frac{\partial{\mathcal L}}{\partial{\bf F}^H} = \sum_{k=1}^{K} \lambda_{{\rm e},k}\left(- {\bf G}_{k}^H{\bf W}_k{\bf B}_k^H{\bf H}_{k}^H + \sum_{j=1}^K{\bf G}_k^H{\bf W}_k{\bf W}_k^H{\bf G}_{k}{\bf F}{\bf H}_j{\bf B}_j{\bf B}_j^H{\bf H}_j^H \right.\\
\left.+ \sigma_{\rm r}^2{\bf G}_k^H{\bf W}_k{\bf W}_k^H{\bf G}_{k}{\bf F}\right) + \lambda_{\rm r}{\bf F} \left(\sum_{k=1}^K{\bf H}_{k}{\bf B}_k{\bf B}_k^H{\bf H}_{k}^H + \sigma_{\rm r}^2{\bf I}_{N_{\rm r}}\right).
\end{multline}
Rearranging the terms in \eqref{dLek}, $\frac{\partial{\mathcal L}}{\partial{\bf F}^H}$ can be expressed as
\begin{multline}\label{dLek2}
\frac{\partial{\mathcal L}}{\partial{\bf F}^H}
=\sum_{k=1}^{K} - \lambda_{{\rm e},k}{\bf G}_{k}^H{\bf W}_k{\bf B}_k^H{\bf H}_{k}^H\\
+ \left(\sum_{i=1}^K\lambda_{{\rm e},i}{\bf G}_i^H{\bf W}_i{\bf W}_i^H{\bf G}_i + \lambda_{\rm r}{\bf I}_{N_{\rm r}}\right){\bf F} \left(\sum_{j=1}^K{\bf H}_j{\bf B}_j{\bf B}_j^H{\bf H}_j^H + \sigma_{\rm r}^2{\bf I}_{N_{\rm r}}\right).
\end{multline}
Equating $\frac{\partial{\mathcal L}}{\partial{\bf F}^*} = 0$, we have the optimal relay filter given by
\begin{equation}
{\bf F} 
=\sum_{k = 1}^K{\bf T}_k{\bf D}_k^H
\end{equation}
with
\begin{equation}
\left\{\begin{aligned}
{\bf T}_k& \triangleq \lambda_{{\rm e},k}\left(\sum_{i=1}^K\lambda_{{\rm e},i}{\bf G}_i^H{\bf W}_i{\bf W}_i^H{\bf G}_i + \lambda_{\rm r}{\bf I}_{N_{\rm r}}\right)^{-1}{\bf G}_{k}^H{\bf W}_k,\\
{\bf D}_k &\triangleq \left(\sum_{j=1}^K{\bf H}_j{\bf B}_j{\bf B}_j^H{\bf H}_j^H+ \sigma_{\rm r}^2{\bf I}_{N_{\rm r}}\right)^{-1}{\bf H}_{k}{\bf B}_k.
\end{aligned}\right.
\end{equation}
Denoting ${\bf T} \triangleq \left[{\bf T}_1 \cdots {\bf T}_K \right]$ and ${\bf D} \triangleq \left[{\bf D}_1 \cdots {\bf D}_K \right]$, ${\bf F}$ can be expressed as ${\bf F} = {\bf T}{\bf D}^H$. \hfill $\Box$

\subsection{Proof of Theorem~\ref{thm_mse}}\label{proof_thm_mse}
The MSE in \eqref{mnx_a} can be rewritten as
\begin{eqnarray}
E_k \!\!\!&=&\!\!\! \left[{\bf I}_{N_{\rm s},k} + {\bf B}_k^H{\bf H}_{k}^H{\bf F}^H{\bf G}_k^H \bar{\bf C}_k^{-1}{\bf G}_{k}{\bf F}{\bf H}_k{\bf B}_k\right]^{-1}\\
\!\!\!&=&\!\!\! {\rm tr}\left({\bf I}_{N_{\rm s},k} - {\bf B}_k^H{\bf H}_{k}^H{\bf F}^H{\bf G}_k^H\left({\bf G}_{k}{\bf F}{\bf H}_k{\bf B}_k{\bf B}_k^H{\bf H}_{k}^H{\bf F}^H{\bf G}_k^H + \bar{\bf C}_k\right)^{-1}{\bf G}_{k}{\bf F}{\bf H}_k{\bf B}_k\right)\nonumber\\
\!\!\!& &\!\!\!\label{eq56}\\
\!\!\!&=&\!\!\! {\rm tr}\left({\bf I}_{N_{\rm s},k} - {\bf B}_k^H{\bf H}_{k}^H{\bf F}^H{\bf G}_k^H\left({\bf G}_{k}{\bf F}{\boldsymbol\Psi}{\bf F}^H{\bf G}_k^H + \sigma_{\rm d}^2{\bf I}_{N_{\rm d},k}\right)^{-1}{\bf G}_{k}{\bf F}{\bf H}_k{\bf B}_k\right)\label{eq57}\\
\!\!\!&=&\!\!\! {\rm tr}\left({\bf I}_{N_{\rm s},k} - {\bf B}_k^H{\bf H}_{k}^H\left[{\boldsymbol\Psi}^{-1} - \left({\boldsymbol\Psi}{\bf F}^H{\bf G}_k^H{\bf G}_{k}{\bf F}{\boldsymbol\Psi} + {\boldsymbol\Psi}\right)^{-1}\right]{\bf H}_k{\bf B}_k\right)\label{eq58}\\
\!\!\!&=&\!\!\! {\rm tr}\left({\bf I}_{N_{\rm s},k} + {\bf B}_k^H{\bf H}_{k}^H{\boldsymbol\Psi}_{\bar k}^{-1}{\bf H}_k{\bf B}_k\right)^{-1} \! + {\rm tr}\left({\bf B}_k^H{\bf H}_{k}^H\left({\boldsymbol\Psi}{\bf F}^H{\bf G}_k^H{\bf G}_{k}{\bf F}{\boldsymbol\Psi} + {\boldsymbol\Psi}\right)^{-1}\right.\nonumber\\
\!\!\!& &\!\!\!\qquad\qquad\qquad\qquad\qquad\qquad\qquad\qquad\qquad\qquad\qquad\qquad\left.\times{\bf H}_k{\bf B}_k\right),\label{sum_ei}
\end{eqnarray}
where we used matrix inversion lemma $\left({\bf A} + {\bf BCD}\right)^{-1}= {\bf A}^{-1} - {\bf A}^{-1}{\bf B}\left({\bf DA}^{-1}{\bf B}\right.\\\left. + {\bf C}^{-1}\right)^{-1}{\bf DA}^{-1}$ to obtain \eqref{eq56} and the first term in \eqref{sum_ei} whereas the matrix identity ${\bf B}^H({\bf B}{\bf C}{\bf B}^H + {\bf I})^{-1}{\bf B} = {\bf C}^{-1} - ({\bf C}{\bf B}^H{\bf B}{\bf C} + {\bf C})^{-1}$ is used to obtain \eqref{eq58} in the above derivation. Note that the first term in \eqref{sum_ei} is irrelevant to ${\bf F}$. Hence for given source matrices, the problem of optimizing ${\bf F}$ can be simplified as
\begin{subequations}
\label{minF}
\begin{eqnarray}
\min_{{\bf F}} \!\!\!& &\!\!\! {\rm tr}\left({\bf B}_k^H{\bf H}_{k}^H\left({\boldsymbol\Psi}{\bf F}^H{\bf G}_k^H{\bf G}_{k}{\bf F}{\boldsymbol\Psi} + {\boldsymbol\Psi}\right)^{-1}{\bf H}_k{\bf B}_k\right) \label{mnF_o}\\
{\rm s.t.} \!\!\!& &\!\!\! {\rm tr}\left({\bf F} {\boldsymbol \Psi} {\bf F}^H\right)\leq P_{{\rm r}}.\label{mnF_c1}
\end{eqnarray}
\end{subequations}
By introducing $\tilde{\bf F} = {\bf F} {\boldsymbol \Psi}^{\frac{1}{2}}$, problem \eqref{minF} can be rewritten as
\begin{subequations}
\label{minF2}
\begin{eqnarray}
\min_{\tilde{\bf F}} \!\!\!& &\!\!\! {\rm tr}\left({\bf B}_k^H{\bf H}_{k}^H{\boldsymbol \Psi}^{-\frac{1}{2}}\left(\tilde{\bf F}^H{\bf G}_k^H{\bf G}_{k}\tilde{\bf F} + {\bf I}_{N_{\rm r}}\right)^{-1}{\boldsymbol \Psi}^{-\frac{1}{2}}{\bf H}_k{\bf B}_k\right) \label{mnF2_o}\\
{\rm s.t.} \!\!\!& &\!\!\! {\rm tr}\left(\tilde{\bf F}\tilde{\bf F}^H\right)\leq P_{{\rm r}}.\label{mnF2_c1}
\end{eqnarray}
\end{subequations}
Let us write the eigenvalue decomposition (EVD) ${\bf G}_k^H{\bf G}_{k} = {\bf V}_{\rm g}{\boldsymbol\Lambda}_{\rm g}{\bf V}_{\rm g}^H$ and the singular value decomposition (SVD) ${\boldsymbol \Psi}^{-\frac{1}{2}}{\bf H}_k{\bf B}_k = {\bf U}_{\psi}{\boldsymbol\Lambda}_{\psi}{\bf V}_{\psi}^H$. The following lemma defines the optimal $\tilde{\bf F}$.

\begin{lemma}\cite[Lemma 2]{rong_simpli}\label{lemm_struc}
For matrices ${\bf A}, \bar{\bf T}, {\bf H}$ of dimensions $m\times n$, $l\times m$, and $k\times l$, respectively, with $k, l, m \ge n$, $r \triangleq {\rm rank} ({\bf H}) \ge n$ and ${\rm rank}(\bar{\bf T} )= n$, the solution to the optimization problem
\begin{subequations}
\begin{eqnarray}
\min_{\bar{\bf T}} \!\!\!& &\!\!\! {\rm tr}\left({\bf A}^H\left(\bar{\bf T}^H{\bf H}^H{\bf H}\bar{\bf T} + {\bf I}_{m}\right)^{-1}{\bf A}\right) \label{mnF22_o}\\
{\rm s.t.} \!\!\!& &\!\!\! {\rm tr}(\bar{\bf T}\bar{\bf T}^H)\leq p,\label{mnF22_c1}
\end{eqnarray}
\end{subequations}
is given by $\bar{\bf T} = \tilde{\bf V}_{\rm h}{\boldsymbol\Lambda}_{\rm T}{\bf U}_{\rm a}^H$ in terms of the SVD of $\bar{\bf T}$. Here ${\bf H} = {\bf U}_{\rm h}{\boldsymbol\Sigma}_{\rm h}{\bf V}_{\rm h}^H$ and ${\bf A} = {\bf U}_{\rm a}{\boldsymbol\Sigma}_{\rm a}{\bf V}_{\rm a}^H$ are the SVDs of ${\bf H}$ and ${\bf A}$, respectively, with the diagonal elements of ${\boldsymbol\Sigma}_{\rm h}$ and ${\boldsymbol\Sigma}_{\rm a}$ sorted in a decreasing order, and $\tilde{\bf V}_{\rm h}$ contains the leftmost $n$ columns of ${\bf V}_{\rm h}$.
\end{lemma}


According to Lemma~\ref{lemm_struc}, the optimal $\tilde{\bf F}$ in \eqref{minF2} has the SVD $\tilde{\bf F} = \tilde{\bf V}_{\rm g}{\boldsymbol\Lambda}_{\rm f}{\bf U}_{\psi}^H$ where $\tilde{\bf V}_{\rm g}$ contains the left-most columns of ${\bf V}_{\rm g}$ corresponding to the non-zero eigenvalues. Then after some simple manipulations, $\tilde{\bf F}$ can be rewritten as $\tilde{\bf F} = \tilde{\bf V}_{\rm g}{\boldsymbol\Lambda}_{\rm f}{\boldsymbol\Lambda}_{\psi}^{-1}{\bf V}_{\psi}^H{\bf V}_{\psi}{\boldsymbol\Lambda}_{\psi}{\bf U}_{\psi}^H = \tilde{\bf T}{\bf B}_k^H{\bf H}_k^H{\boldsymbol \Psi}^{-\frac{1}{2}}$ where $\tilde{\bf T} \triangleq \tilde{\bf V}_{\rm g}{\boldsymbol\Lambda}_{\rm f}{\boldsymbol\Lambda}_{\psi}^{-1}{\bf V}_{\psi}^H$. Hence ${\bf F}$ can be expressed as ${\bf F} = \tilde{\bf T}{\bf B}_k^H{\bf H}_k^H{\boldsymbol \Psi}^{-1}$. Interestingly, ${\bf F} = \tilde{\bf T}{\bf B}_k^H{\bf H}_k^H{\boldsymbol \Psi}^{-1}$ can be expressed as ${\bf F} = \tilde{\bf T}\tilde{\bf D}^H$, which is structurally identical to the one defined in Theorem~\ref{thm_struc}.

Applying this structure of the relay matrix, the second term in \eqref{sum_ei} can be written as
\begin{eqnarray}
{\rm tr} \!\!\!\!\!\!& &\!\!\!\!\!\! \left({\bf B}_k^H{\bf H}_{k}^H\left({\boldsymbol\Psi}{\bf F}^H{\bf G}_k^H{\bf G}_{k}{\bf F}{\boldsymbol\Psi} + {\boldsymbol\Psi}\right)^{-1}{\bf H}_k{\bf B}_k\right)\nonumber\\
\!\!\!&=&\!\!\! {\rm tr} \left({\bf B}_k^H{\bf H}_{k}^H\left({\boldsymbol\Psi}{\boldsymbol\Psi}^{-1}{\bf H}_k{\bf B}_k\tilde{\bf T}^H{\bf G}_k^H{\bf G}_{k}\tilde{\bf T}{\bf B}_k^H{\bf H}_{k}^H{\boldsymbol\Psi}^{-1}{\boldsymbol\Psi} + {\boldsymbol\Psi}\right)^{-1}{\bf H}_k{\bf B}_k\right)\nonumber\\
\!\!\!&=&\!\!\! {\rm tr} \left({\bf B}_k^H{\bf H}_{k}^H\left({\boldsymbol\Psi}^{-1} - {\boldsymbol\Psi}^{-1}{\bf H}_k{\bf B}_k\left({\bf B}_k^H{\bf H}_{k}^H{\boldsymbol\Psi}^{-1}{\bf H}_k{\bf B}_k + \left(\tilde{\bf T}^H{\bf G}_k^H{\bf G}_{k}\tilde{\bf T}\right)^{-1}\right)^{-1}\right.\right.\nonumber\\
\!\!\!& &\!\!\! \qquad\qquad\qquad\qquad\qquad\qquad\qquad\qquad\qquad\qquad \times\left.\left.{\bf B}_k^H{\bf H}_{k}^H{\boldsymbol\Psi}^{-1}\right){\bf H}_k{\bf B}_k\right)\nonumber\\
\!\!\!&=&\!\!\! {\rm tr} \left({\bf B}_k^H{\bf H}_{k}^H{\boldsymbol\Psi}^{-1}{\bf H}_k{\bf B}_k - {\bf B}_k^H{\bf H}_{k}^H{\boldsymbol\Psi}^{-1}{\bf H}_k{\bf B}_k\left({\bf B}_k^H{\bf H}_{k}^H{\boldsymbol\Psi}^{-1}{\bf H}_k{\bf B}_k\right.\right.\nonumber\\
\!\!\!& &\!\!\! \qquad\qquad\qquad\qquad\qquad\qquad\left.\left. + \left(\tilde{\bf T}^H{\bf G}_k^H{\bf G}_{k}\tilde{\bf T}\right)^{-1}\right)^{-1}{\bf B}_k^H{\bf H}_{k}^H{\boldsymbol\Psi}^{-1}{\bf H}_k{\bf B}_k\right)\nonumber\\
\!\!\!&=&\!\!\! {\rm tr} \left(\left({\bf B}_k^H{\bf H}_{k}^H{\boldsymbol\Psi}^{-1}{\bf H}_k{\bf B}_k\right)^{-1} + \tilde{\bf T}^H{\bf G}_k^H{\bf G}_{k}\tilde{\bf T}\right)^{-1}.
\end{eqnarray}
Thus the MSE in \eqref{mnx_a} can be expressed as the sum of two MSEs given by
\begin{multline}
E_k ={\rm tr}\left({\bf I}_{N_{\rm s},k} + {\bf B}_k^H{\bf H}_{k}^H{\boldsymbol\Psi}_{\bar k}^{-1}{\bf H}_k{\bf B}_k\right)^{-1}\\
 + {\rm tr} \left(\left({\bf B}_k^H{\bf H}_{k}^H{\boldsymbol\Psi}^{-1}{\bf H}_k{\bf B}_k\right)^{-1} + \tilde{\bf T}^H{\bf G}_k^H{\bf G}_{k}\tilde{\bf T}\right)^{-1}.
\end{multline}\hfill$\Box$

\subsection{Proof of Proposition~\ref{prop_approx}}\label{proof_prop_approx}
Assuming that the first-hop SNR is reasonably high, it emerges that $\sum_{j=1}^K{\bf H}_j{\bf B}_j{\bf B}_j^H{\bf H}_j^H\\ \gg \sigma_{\rm r}^2{\bf I}_{N_{\rm r}}$ where ${\bf A}\gg {\bf B}$ effectively means that the eigenvalues of ${\bf A} - {\bf B}$ are much greater than zero. Hence,
\begin{eqnarray}
{\bf B}_k^H{\bf H}_{k}^H{\boldsymbol\Psi}^{-1}{\bf H}_k{\bf B}_k \!\!\!&=&\!\!\! {\bf B}_k^H{\bf H}_{k}^H\left(\sum_{j=1}^K{\bf H}_j{\bf B}_j{\bf B}_j^H{\bf H}_j^H + \sigma_{\rm r}^2{\bf I}_{N_{\rm r}}\right)^{-1}{\bf H}_k{\bf B}_k\nonumber\\
\!\!\!&\approx&\!\!\! {\bf B}_k^H{\bf H}_{k}^H\left(\sum_{j=1}^K{\bf H}_j{\bf B}_j{\bf B}_j^H{\bf H}_j^H\right)^{-1}{\bf H}_k{\bf B}_k.\label{bPsi}
\end{eqnarray}
Let ${\bf U}_k{\boldsymbol \Lambda}_k{\bf U}_k^H$ be the EVD of ${\bf H}_k{\bf B}_k{\bf B}_k^H{\bf H}_k^H$. Without loss of generality, we express ${\bf U}_k = \left[{\bf U}_k^{(\bar 0)} \,\,{\bf U}_k^{(0)}\right]$ and ${\boldsymbol \Lambda}_k = \left[\begin{array}{cc}
{\boldsymbol \Lambda}_k^{(\bar 0)}\,\, & {\bf 0}\\
{\bf 0} & {\bf 0}
\end{array}
\right]$, where ${\bf U}_k^{(\bar 0)}$ and ${\bf U}_k^{(0)}$ contain the eigenvectors corresponding to the non-zero and zero eigenvalues, respectively, in ${\bf U}_k$ while ${\boldsymbol \Lambda}_k^{(\bar 0)}$ is an $N_{{\rm b},k}\times N_{{\rm b},k}$ diagonal matrix containing the non-zero eigenvalues as the main diagonal. Thus ${\bf H}_k{\bf B}_k = {\bf U}_k\bar{\boldsymbol \Lambda}_k^{(\bar 0)}$ where $\bar{\boldsymbol \Lambda}_k^{(\bar 0)} = \left[\begin{array}{c}
{\boldsymbol \Lambda}_k^{(\bar 0)\frac{1}{2}}\\
{\bf 0}
\end{array}
\right]$. Similarly, we obtain the following EVD
\begin{eqnarray}
\sum_{j=1\atop k\neq k}^K{\bf H}_j{\bf B}_j{\bf B}_j^H{\bf H}_j^H = {\bf U}_{\bar k}{\boldsymbol \Lambda}_{\bar k}{\bf U}_{\bar k}^H
&=\left[{\bf U}_{\bar k}^{(\bar 0)} \,\,{\bf U}_{\bar k}^{(0)}\right] \left[\begin{array}{cc}
{\boldsymbol \Lambda}_{\bar k}^{(\bar 0)}\,\, & {\bf 0}\\
{\bf 0} & {\bf 0}
\end{array}
\right] \left[{\bf U}_{\bar k}^{(\bar 0)} \,\,{\bf U}_{\bar k}^{(0)}\right]^H\\
&=\left[{\bf U}_{\bar k}^{(0)} \,\,{\bf U}_{\bar k}^{(\bar 0)}\right] \left[\begin{array}{cc}
{\bf 0}\,\, & {\bf 0}\\
{\bf 0} & {\boldsymbol \Lambda}_{\bar k}^{(\bar 0)}
\end{array}
\right] \left[{\bf U}_{\bar k}^{(0)} \,\,{\bf U}_{\bar k}^{(\bar 0)}\right]^H.
\end{eqnarray}

Substituting ${\bf H}_k{\bf B}_k$ in \eqref{bPsi} with ${\bf H}_k{\bf B}_k = {\bf U}_k\bar{\boldsymbol \Lambda}_k^{(\bar 0)}$, we obtain
\begin{eqnarray}
{\bf B}_k^H{\bf H}_{k}^H\left(\sum_{j=1}^K{\bf H}_j{\bf B}_j{\bf B}_j^H{\bf H}_j^H\right)^{-1}{\bf H}_k{\bf B}_k \!\!\!&=&\!\!\! \bar{\boldsymbol \Lambda}_k^{(\bar 0)H}\left({\boldsymbol \Lambda}_k + {\bf U}_k^H{\bf U}_{\bar k}{\boldsymbol \Lambda}_{\bar k}{\bf U}_{\bar k}^H{\bf U}_k\right)^{-1}\bar{\boldsymbol \Lambda}_k^{(\bar 0)}.\nonumber\\
\label{bPsi_eq}
\end{eqnarray}
Now we rewrite ${\bf U}_k^H{\bf U}_{\bar k}$ as
\begin{eqnarray}
{\bf U}_k^H{\bf U}_{\bar k} = \left[{\bf U}_k^{(\bar 0)} \,\,{\bf U}_k^{(0)}\right]^H\left[{\bf U}_{\bar k}^{(0)} \,\,{\bf U}_{\bar k}^{(\bar 0)}\right] = \left[\begin{array}{cc}
\bar{\bf U}_k^{(0)}\,\, & {\bf 0}\\
{\bf 0} & \bar{\bf U}_k^{(\bar 0)}
\end{array}
\right],
\end{eqnarray}
where $\bar{\bf U}_k^{(0)}$ and $\bar{\bf U}_k^{(\bar 0)}$ are $N_{{\rm b},k} \times N_{{\rm b},k}$ and $\left(N_{\rm r}-N_{{\rm b},k}\right)\times \left(N_{\rm r}-N_{{\rm b},k}\right)$ unitary matrices, respectively. As a consequence, we obtain
\begin{multline}
{\bf U}_k^H{\bf U}_{\bar k}{\boldsymbol \Lambda}_{\bar k}{\bf U}_{\bar k}^H{\bf U}_k = {\bf U}_k^H\left[{\bf U}_{\bar k}^{(0)} \,\,{\bf U}_{\bar k}^{(\bar 0)}\right] \left[\begin{array}{cc}
{\bf 0}\,\, & {\bf 0}\\
{\bf 0} & {\boldsymbol \Lambda}_{\bar k}^{(\bar 0)}
\end{array}
\right] \left[{\bf U}_{\bar k}^{(0)} \,\,{\bf U}_{\bar k}^{(\bar 0)}\right]^H{\bf U}_k\\
 = \left[\begin{array}{cc}
{\bf 0}\,\, & {\bf 0}\\
{\bf 0} & \bar{\bf U}_k^{(\bar 0)}{\boldsymbol \Lambda}_{\bar k}^{(\bar 0)}\bar{\bf U}_k^{(\bar 0)H}
\end{array}
\right].
\end{multline}
Using the identity ${\bf U}^{-1} = {\bf U}^H$ for a unitary matrix ${\bf U}$, we obtain
\begin{eqnarray}
\left({\boldsymbol \Lambda}_k + {\bf U}_k^H{\bf U}_{\bar k}{\boldsymbol \Lambda}_{\bar k}{\bf U}_{\bar k}^H{\bf U}_k\right)^{-1} = \left[\begin{array}{cc}
{\boldsymbol \Lambda}_k^{(\bar 0)^{-1}}\,\, & {\bf 0}\\
{\bf 0} & \bar{\bf U}_k^{(\bar 0)}{\boldsymbol \Lambda}_{\bar k}^{(\bar 0)^{-1}}\bar{\bf U}_k^{(\bar 0)H}
\end{array}
\right].\label{Rii}
\end{eqnarray}
Substituting \eqref{Rii} into \eqref{bPsi_eq}, we obtain
\begin{multline}\label{bPsi_eq2}
{\bf B}_k^H{\bf H}_{k}^H\left(\sum_{j=1}^K{\bf H}_j{\bf B}_j{\bf B}_j^H{\bf H}_j^H\right)^{-1}{\bf H}_k{\bf B}_k\\
 = \left[\begin{array}{cc}
{\boldsymbol \Lambda}_k^{(\bar 0)\frac{1}{2}H} &
{\bf 0}
\end{array}
\right] \left[\begin{array}{cc}
{\boldsymbol \Lambda}_k^{(\bar 0)^{-1}}\,\, & {\bf 0}\\
{\bf 0} & \bar{\bf U}_k^{(\bar 0)}{\boldsymbol \Lambda}_{\bar k}^{(\bar 0)^{-1}}\bar{\bf U}_k^{(\bar 0)H}
\end{array}
\right]
\left[\begin{array}{c}
{\boldsymbol \Lambda}_k^{(\bar 0)\frac{1}{2}}\\
{\bf 0}
\end{array}
\right]\\
={\boldsymbol \Lambda}_k^{(\bar 0)\frac{1}{2}H}{\boldsymbol \Lambda}_k^{(\bar 0)^{-1}}{\boldsymbol \Lambda}_k^{(\bar 0)\frac{1}{2}}={\bf I}_{N_{{\rm b},k}}.
\end{multline}
Thus for high first-hop SNR, ${\bf B}_k^H{\bf H}_{k}^H{\boldsymbol\Psi}^{-1}{\bf H}_k{\bf B}_k$ can be approximated as ${\bf I}_{N_{{\rm b},k}}$.\hfill$\Box$


\begin{backmatter}

\section*{Competing interests}
  The authors declare that they have no competing interests.

\section*{Funding}
  This work is supported by EPSRC under grant EP/K015893/1.




\bibliographystyle{bmc-mathphys} 
\bibliography{refdb, mypubdb}      




\section*{Figures}
\begin{figure}[h!]
\centering
\includegraphics*[scale=.9]{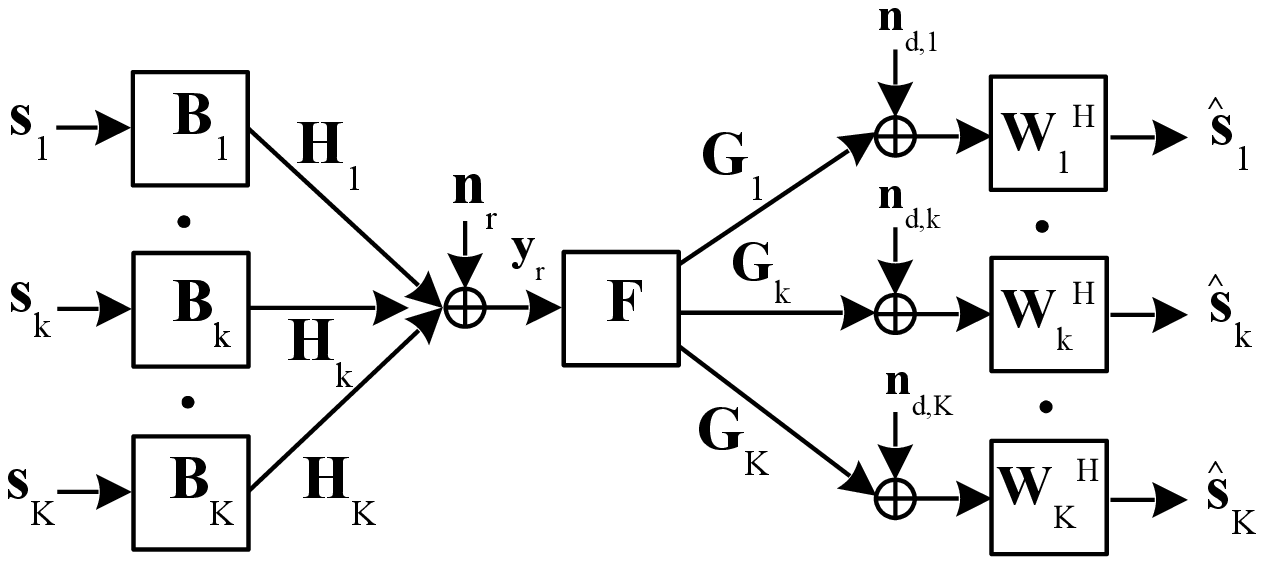}
\caption{The model of the dual-hop interference MIMO relay system.}
\label{sysmod}
\end{figure}

\begin{figure}[h!]
\centering{\includegraphics*[scale=.9]{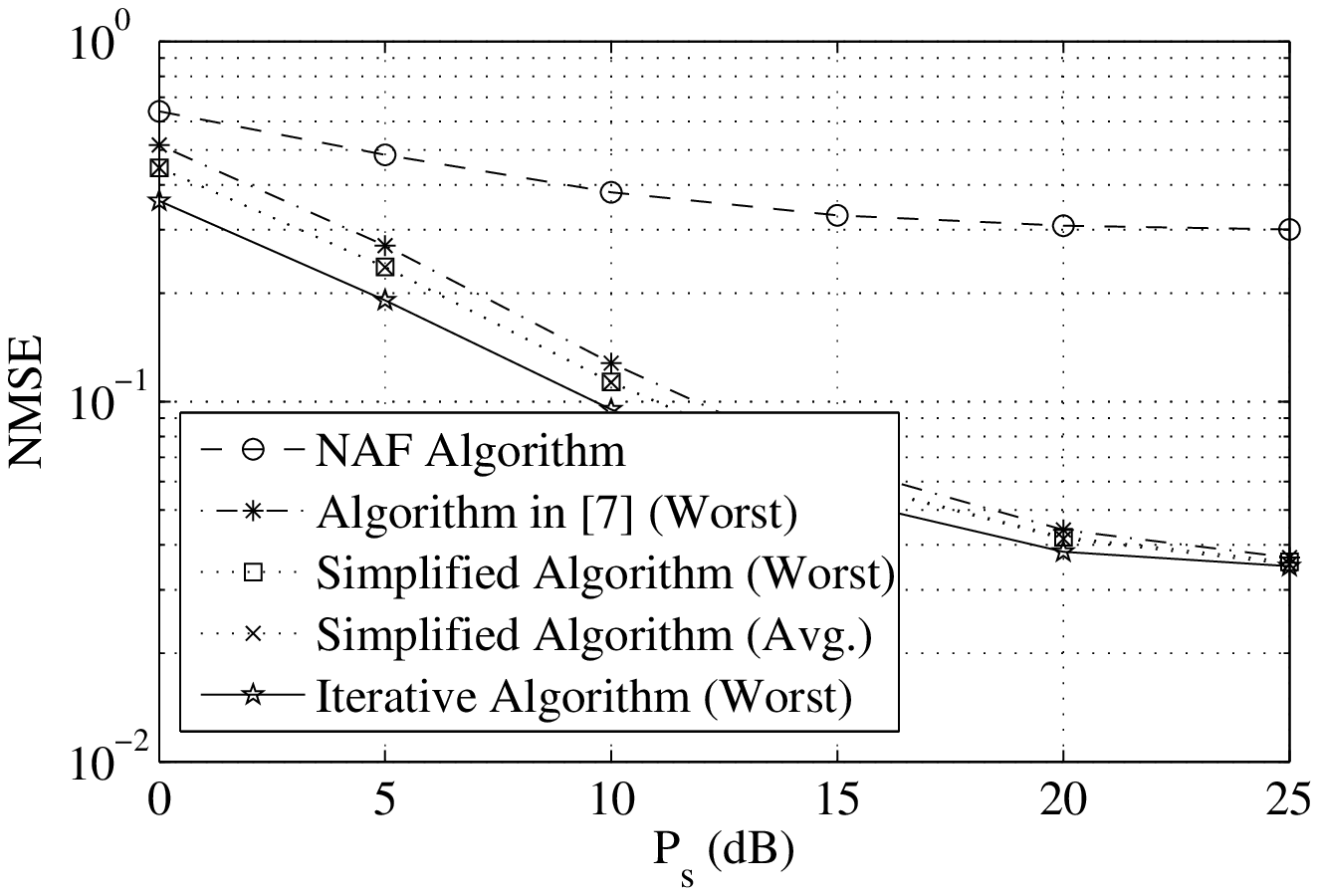}}
\caption{Example 1: Normalized MSE versus $P_{\rm s}$.
{$K = 3$, $N_{\rm s} = 3, N_{\rm r} = 9$, $N_{\rm d} = 3$,} $P_{\rm r} = 20$dB.}
\label{mseL4s}
\end{figure}

\begin{figure}[h!]
\centering
\includegraphics*[scale=.9]{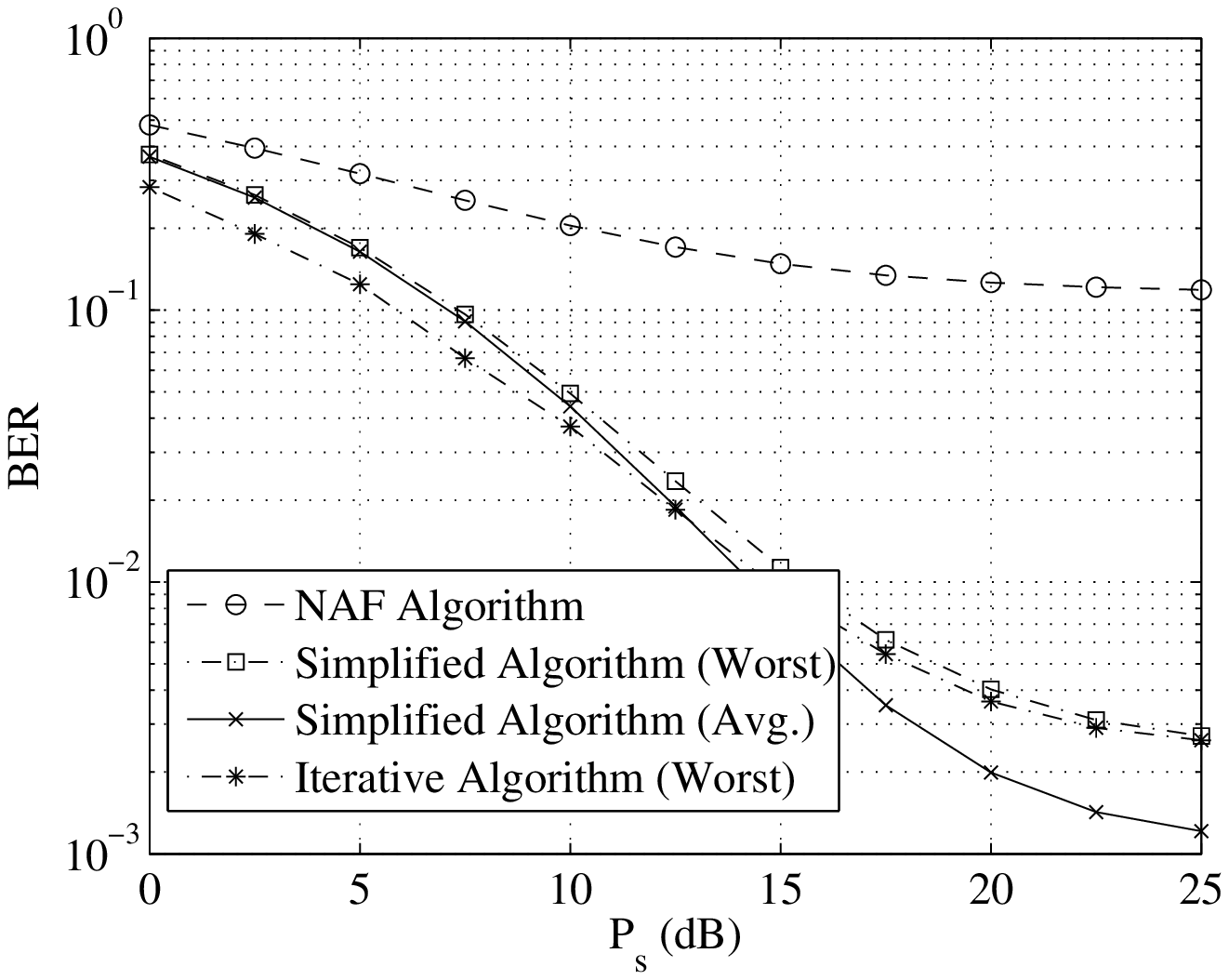} 
\caption{Example 2: BER versus $P_{\rm s}$.
$K = 3$, $N_{\rm s} = 2, N_{\rm r} = 6$, $N_{\rm d} = 3$, $P_{\rm r} = 20$dB.}
\label{berL2s}
\end{figure}

\begin{figure}[h!]
\centering
\includegraphics*[scale=.9]{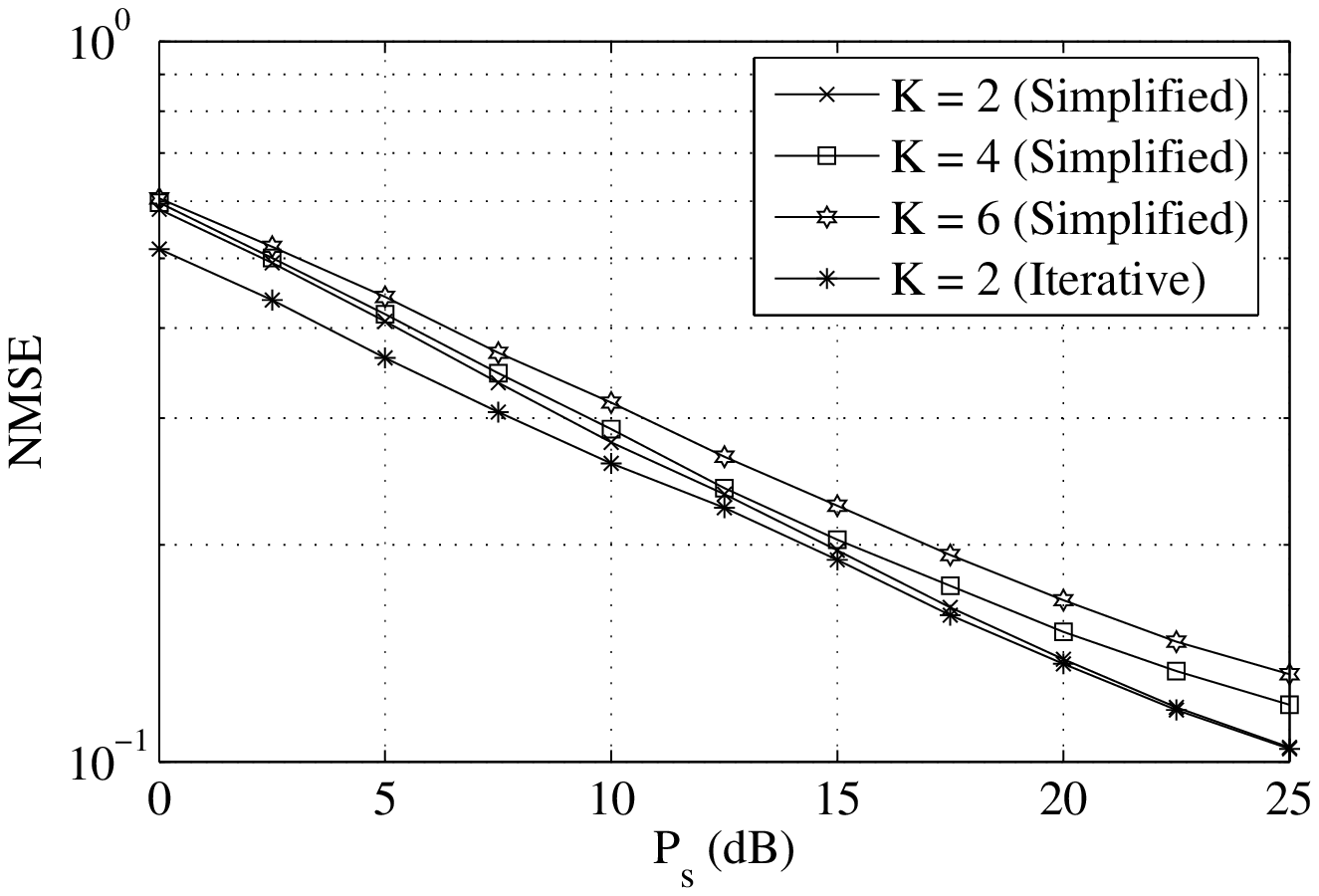} 
\caption{Example 3: MSE versus $P_{\rm s}$ in two-way relaying. Varying number of links,
$N_{\rm s} = 2, N_{\rm r} = K N_{\rm s}$, $N_{\rm d} = 6$, $P_{\rm r} = 20$dB.}
\label{2way_diffKs}
\end{figure}

\begin{figure}[h!]
\centering
\includegraphics*[scale=.9]{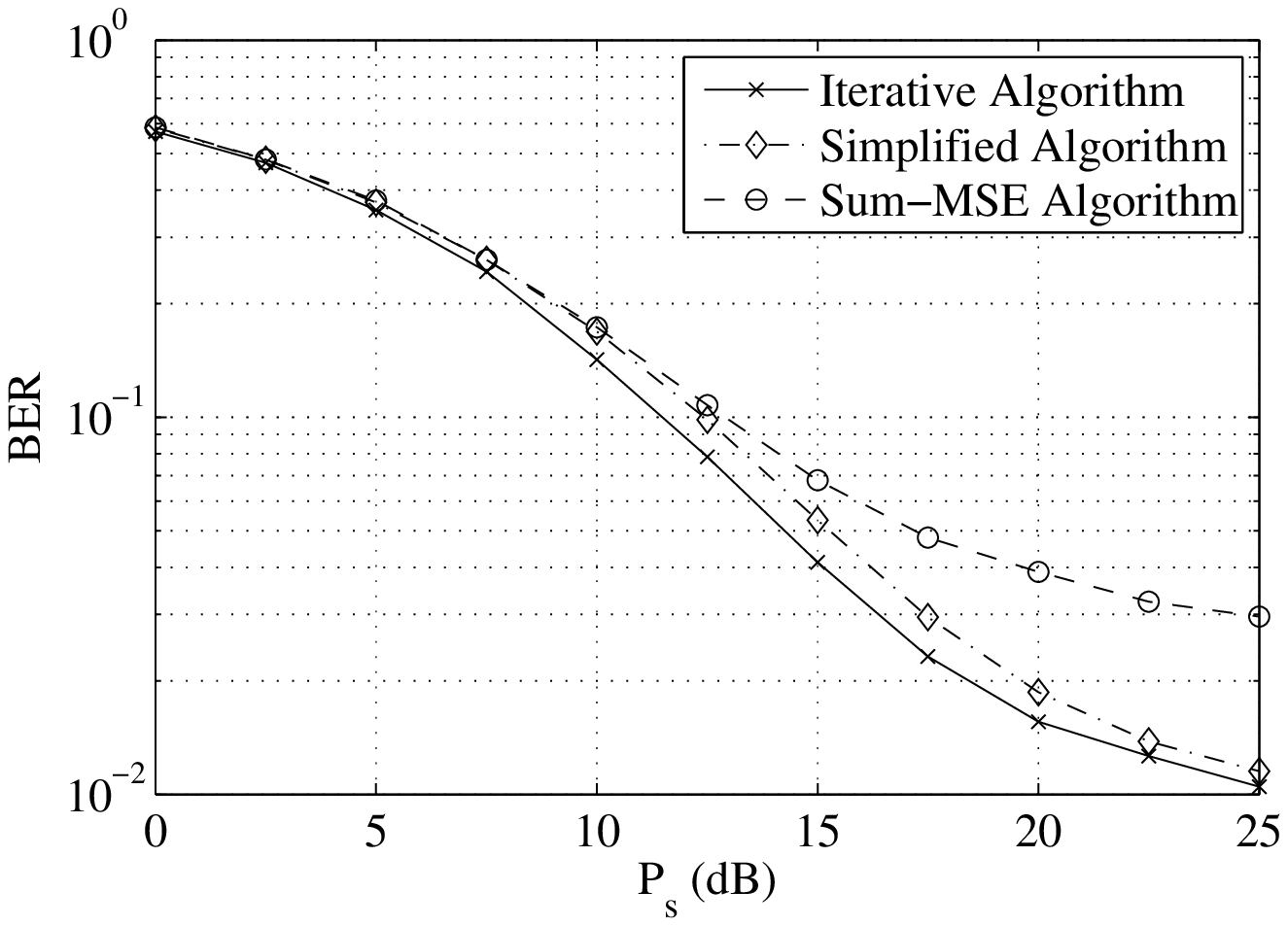} 
\caption{Example 4: BER versus $P_{\rm s}$ in two-way relaying for different algorithms,
$N_{\rm s} = 2, K = 3, N_{\rm r} = K N_{\rm s}$, $N_{\rm d} = 6$, $P_{\rm r} = 20$dB.}
\label{2way_diffAlgs}
\end{figure}


\section*{Tables}

\begin{table}[h!]
\noindent\makebox[\linewidth]{\rule{\textwidth}{0.5pt}}
\caption{Iterative solution of problem (\ref{minmax0})}\label{tab_it}
\noindent\makebox[\linewidth]{\rule{\textwidth}{0.5pt}}
\begin{enumerate}
\item Randomly initialize ${\bf F}$ and $\{{\bf B}_k\}$ such that the constraints (\ref{mnx_a_c1}) and (\ref{mnx_a_c2}) are satisfied.
\item Repeat
\begin{enumerate}
\item Obtain $\{{\bf W}_k\}$ as defined in (\ref{Wk}) using known $\{{\bf B}_k\}$ and ${\bf F}$.
\item Solve the subproblem (\ref{minmaxF2}) to update ${\bf F}$ using fixed $\{{\bf W}_k\}$ and $\{{\bf B}_k\}$.
\item Update $\{{\bf B}_k\}$ through solving the subproblem (\ref{sdp_itb}) using ${\bf F}$ and $\{{\bf W}_k\}$ known from the previous steps.
\end{enumerate}
\item Until convergence.
\end{enumerate}
\noindent\makebox[\linewidth]{\rule{\textwidth}{0.5pt}}
\end{table}

\begin{table}[h!]
\noindent\makebox[\linewidth]{\rule{\textwidth}{0.5pt}}
\caption{Proposed simplified algorithm for solving problem (\ref{minmax0})}\label{tab_simpli}
\noindent\makebox[\linewidth]{\rule{\textwidth}{0.5pt}}
\begin{enumerate}
\item Initialize ${\bf B}_k, \forall k,$ satisfying the constraints \eqref{decs2_c2}.
\item Repeat
    \begin{enumerate}
    \item Update ${\bf D}_k, \forall k,$ as in (\ref{Dk}).
    \item Update ${\bf B}_k, \forall k$, through solving the subproblem \eqref{decs2}.
    \end{enumerate}
\item Until convergence.
\item Solve the subproblem \eqref{decr3} to obtain ${\bf Q}$.
\end{enumerate}
\noindent\makebox[\linewidth]{\rule{\textwidth}{0.5pt}}
\end{table}

\begin{table}[h!]
\centering \caption{Iterations required till convergence in the
proposed algorithm} \label{tab2}
\begin{tabular}[t]{|c|c|c|c|c|c|c|}
\hline
$P_{\rm s}$ (dB) & $0$ & $5$ & $10$ & $15$ & $20$ & $25$\\
\hline
Iterations & $3$ & $3$ & $3$ & $4$ & $5$ & $5$\\
\hline
\end{tabular}
\end{table}


%

\end{backmatter}
\end{document}